\documentclass[nofootinbib,floatfix,superscriptaddress,twocolumn]{revtex4}        
\usepackage{graphicx}
\usepackage{epsfig}
\usepackage{bm}
\usepackage[T1]{fontenc}
\usepackage[latin9]{inputenc}
\usepackage{amssymb}
\usepackage{float}
\usepackage{amsmath}
\usepackage{dcolumn}
\usepackage{cancel}
\usepackage[colorlinks]{hyperref}
\usepackage[usenames,dvipsnames]{color}
\hypersetup{
     breaklinks=true,
    pdfstartview={FitH},    
    colorlinks=true,       
    linkcolor=blue,          
    citecolor=red,        
    filecolor=magenta,      
    urlcolor=blue,           
    anchorcolor=green,      
    linktocpage=true
}

\newcommand{\be}{\begin{equation}}
\newcommand{\ee}{\end{equation}}

\begin{document}

\title{Saez-Ballester gravity in Kantowski-Sachs Universe: a new reconstruction paradigm for Barrow Holographic Dark Energy}

\author{G.~G.~Luciano}
\email{giuseppegaetano.luciano@udl.cat}
\affiliation{Applied Physics Section of Environmental Science Department, Escola Polit\`ecnica Superior, Universitat de Lleida, Av. Jaume
II, 69, 25001 Lleida, Spain}

\date{\today}

\begin{abstract}
We reconstruct Barrow Holographic Dark Energy (BHDE)
within the framework of Saez-Ballester Scalar Tensor Theory.
As a specific background, we consider 
a homogeneous and anisotropic Kantowski-Sachs
Universe filled up with BHDE and dark matter. 
By assuming the Hubble radius as an IR cutoff, we
investigate both the cases of non-interacting and interacting
dark energy scenarios. We analyze the evolutionary behavior
of various model parameters, 
such as skewness parameter, Equation-of-State parameter, 
deceleration parameter, jerk parameter and
squared sound speed. We also draw the trajectories
of $\omega_D-\omega'_D$ phase plane and examine statefinder diagnosis. Observational consistency is discussed
by inferring the current value of Hubble's parameter through the
best fit curve of data points from Differential Age (DA) and Baryon Acoustic Oscillations (BAO) and commenting on cosmological perturbations and growth rate of matter fluctuations in BHDE.
We show that our model satisfactorily retraces
the history of the Universe, thus providing a potential candidate explanation for dark energy. 
Comparison with other reconstructions of BHDE is finally analyzed. 
\end{abstract}

 \maketitle

\section{Introduction}
\label{Intro}
Precision Cosmology~\cite{Primack:2006it}
measurements have definitively shown
that our Universe is experiencing an accelerated phase expansion~\cite{Supern,Supernbis,Supernter,Supernquar,Supernquin,ER,Vag1,Vag2}. 
However, the fuel of this mechanism is not yet known, leaving room for disparate explanations. Tentative descriptions
can be basically grouped into two classes:
on one side, Extended Gravity Theories~\cite{CapozDela}  
aim at solving the puzzle by modifying the geometric
part of Einstein-Hilbert action in General Relativity.
On the other side, 
one can introduce new degrees of freedom (DoF) in the matter
sector, giving rise to dynamical Dark Energy models. 
In this context, a largely followed 
approach is the so-called Holographic Dark Energy (HDE) model~\cite{Cohen:1998zx,Horava:2000tb,Thomas:2002pq,Li:2004rb,Hsu:2004ri,Huang:2004ai,Nojiri:2005pu,Wang:2005ph,Setare:2006sv,Guberina:2006qh,Granda,Sheykhi:2011cn,Bamba:2012cp,Ghaffari:2014pxa,Wang:2016och,Odi1,
Moradpour:2020dfm,Zhang,Li,Zhang2,Lu,Nojiri:2019kkp}, 
which is based on the use of the holographic principle
at cosmological scales. 

In the lines of gravity-thermodynamic conjecture, HDE 
describes our Universe as a hologram, the DoF of which
are encoded by Bekenstein-Hawking entropy. 
Nevertheless, it fails to retrace the evolution of the cosmos properly~\cite{Horava:2000tb,Thomas:2002pq,Li:2004rb}, thus
motivating suitable amendments to be implemented. 
Along this line, a promising framework is offered by
HDE with deformed horizon entropies~\cite{Odi,Odi2}, such as Tsallis~\cite{Tsallis1,Tsallis2,Tsallis3,Tsallis4}, Kaniadakis~\cite{Kana1,Kana2,Kana3} and Barrow~\cite{Bar1,Bar2,Bar3,Bar4,Bar5,Bar6,Bar7,Lucianoarx,BarUlt} entropies, 
which arise from the effort to 
introduce non-extensive, relativistic and quantum gravity
corrections in the classical Boltzmann-Gibbs statistics, respectively. 
While predicting a richer phenomenology comparing to the standard Cosmology, generalized HDE models suffer from the absence of an underlying Lagrangian. This somehow questions their relevance in improving our knowledge
of Universe at fundamental level.

Preliminary attempts to overcome the above issue have been made by considering reconstructing scenarios, where
effective Lagrangian models are built by comparing 
extended HDE and modified gravity. 
So far, this recipe has extensively been 
used for Tsallis HDE, with a number of 
applications in $f(R)$~\cite{EPL}, $f(R,T)$~\cite{Chinese}, 
$f(G,T)$~\cite{fgtgrav}, teleparallel~\cite{Wahe}, Brans-Dicke~\cite{GhaffariBD}, logarithmic Brans-Dicke~\cite{LogBD}
and tachyon~\cite{LiuT} models, among others.
By contrast, comparably less attention has been
devoted to Barrow HDE~\cite{Sarkar,LucianoPRD,TeleBar}.
However, it is such a framework that
can potentially open new perspectives in modern theoretical Cosmology, especially in light of the quantum gravitational nature
of the underlying Barrow's conjecture~\cite{B2020}. And, in fact, 
in the absence of any fully quantum theory of Universe, the best we can do toward formulating the quantum effective action of cosmological model at this stage is to frame common inputs and empirical predictions of existing quantum gravity-oriented cosmological models in suitable modifications of General Relativity.

Among the several modifications of Einstein's theory~\cite{SB}, 
Saez-Ballester Theory (SBT) has recently proven to be 
versatile enough to both address the dark energy problem
and accommodate reconstructing scenarios~\cite{SBTBianchi1,Rasouli1,SBTBianchi2,Rasouli2,Santhi}.
SBT is a member of the class of Scalar Tensor Theory of gravity.
In this theory the metric potentials are coupled to
a scalar field, which notoriously plays a key role 
in gravitation and cosmology (see~\cite{Kim,Guth,Linde}). 
In a broader context, SBT has been discussed in 
Bianchi Cosmology in~\cite{SBTBianchi1,SBTBianchi2}, 
reproducing the transition from decelerating Universe
to accelerating phase. On the other hand, in~\cite{Santhi}
it has been considered as a background to 
investigate Tsallis HDE. The ensuing model 
exhibits qualitative consistency with observations, 
though it is classically unstable. 
Currently, SBT and, in general, Extended Gravity
are held to align with Precision Cosmology data.

Starting from the above premises, in this work
we propose a reconstruction of BHDE in SBT.  
We frame our analysis in Kantowski-Sachs (KS) 
geometry~\cite{KSM}, which
describes a homogeneous but anisotropic Universe, the spatial section of which has the topology of $\mathbb{R}\times S^2$ (see also~\cite{Req1} for a recent study of the evolution of KS Universe towards de Sitter at late time). 
The reason why we consider such a type of Universe
is that theoretical investigations and new probes such as Cosmic Background Explorer (COBE), Wilkinson Microwave Anisotropy Probe (WMAP) and Planck have recorded the presence of anisotropy in our Universe~\cite{Ani0,Ani1}, thus requiring a generalization of the canonical
Friedmann-Lemaitre-Robertson-Walker model. 
This is also confirmed by
recent space-based X-ray observations of hundreds of galaxy clusters~\cite{Ani}. 
Motivated by these arguments, we explore 
the history of a KS Universe filled up with anisotropic BHDE
and dark matter in SBT. We construct both non-interacting and interacting
models by assuming the Hubble radius
as an IR cutoff and
solving the field equations for a particular
relationship between the metric potentials.
We focus on the evaluation 
of skewness parameter, Equation-of-State parameter, 
deceleration parameter, jerk parameter and 
squared sound speed. Also, we draw the trajectories
of $\omega_D-\omega'_D$ phase plane and discuss
the statefinder diagnosis. {We discuss observational consistency by deriving the current value of Hubble's parameter through the best fit curve of 57 data points measured from Differential Age (DA) and Baryon Acoustic Oscillations (BAO) and commenting on cosmological perturbations and structure formation.}
We show that our model explains
the current expansion satisfactorily, thus
providing a potential candidate for dark energy. 
Comparison with observations enables us
to constrain the values of free parameters in SBT.

The remainder of the work is organized as follows:
in the next Section we review the basics of BHDE and SBT 
in Kantowski-Sachs Universe.  
To this aim, we follow~\cite{Santhi}.
Section~\ref{reco} is devoted to analyze the cosmic history of reconstructed BHDE, while in Sec.~\ref{Obco} we discuss
Hubble's parameter evolution and cosmological perturbations. 
Conclusions and outlook are finally summarized in Sec.~\ref{C&O}. 
Throughout the whole manuscript, we use natural units. 

\section{Saez-Ballester Theory and Barrow Holographic Dark Energy: An Introduction}
\label{SBg}

In this Section we set the notation and provide
the basic ingredients for the core analysis of this work. We 
first define the geometry of a KS Universe in SBT, 
deriving the corresponding field equations. Then, 
we focus on BHDE framework and its main advantages over the  standard HDE scenario. In passing, we mention that
a similar study has recently been performed
in~\cite{Bianchi_I} in Bianchi-I anisotropic Universe with BHDE.

\subsection{Saez-Ballester Theory of Gravity}

In SB Scalar Tensor Theory of gravity
the Lagrangian is written in the form~\cite{SB}
\be
\mathcal{L}_{SB}\,=\,R-w\phi^n\phi,_\gamma\phi^{,\gamma}\,,
\ee
where $R$ is the scalar curvature, $\phi$ a dimensionless
scalar field, $w$ and $n$ arbitrary dimensionless constants
and $\phi^{,\gamma}\equiv\phi,_\alpha g^{\alpha\gamma}$ 
(as usual we denote partial derivatives by a comma, 
while covariant derivates by a semicolon).  

From the above Lagrangian, one can build the action
\be
I_{SB}\,=\,\int_{\sigma}(\mathcal{L}_{SB}+\mathcal{L}_m)(-g)^{1/2}\,dX^1dX^2dX^3dX^4\,,
\ee
up to an overall  factor multiplying 
the matter Lagrangian $\mathcal{L}_m$. Here
$g$ is the determinant of the metric, $X^i$ the coordinates and 
$\Sigma$ an arbitrary domain of integration. 

Now, for arbitrary independent variations of the metric
and the scalar field vanishing at the boundary surface of
$\Sigma$, the variational principle $\delta I_{SB}=0$ leads
to the field equations
\begin{eqnarray}
\label{F1}
\hspace{-3mm}G_{\alpha\beta}-w \phi^n\left(\phi,_{\alpha}\phi,_\beta-\frac{1}{2}g_{\alpha\beta}\phi,_\gamma\phi^{,\gamma}\right)&\hspace{-1mm}=\hspace{-1mm}&-(T_{\alpha\beta}\,+\,\widetilde T_{\alpha\beta})\,,\\[2mm]
\hspace{-3mm}2\phi^n \phi^{,\gamma}_{;\gamma}\,+\,n\phi^{n-1}\phi,_\gamma\phi^{,\gamma}&\hspace{-1mm}=\hspace{-1mm}&0\,,
\label{F2}
\end{eqnarray}
where $G_{\alpha\beta}$ is the Einstein tensor and $T_{\alpha\beta}$
the energy-momentum tensor (EMT) defined form $\mathcal{L}_m$ in the usual way. For later convenience, here we have separated out the
contribution $\widetilde T_{\alpha\beta}$ due to dark energy.

From relations~\eqref{F1} and~\eqref{F2}, it is easy to prove that the following conservation equation holds
\be
\label{cont}
(T^{\alpha\beta}+\widetilde{T}^{\alpha\beta})_{;\alpha}\,=\,0\,.
\ee
For dark matter of density $\rho_m$ and anisotropic DE of density $\rho_D$, the EMTs read~\cite{Santhi}
\begin{eqnarray}
T_{\alpha\beta}&=&\mathrm{diag}\left[1,0,0,0\right]\rho_m\,,\\[2mm]
\nonumber
\widetilde T_{\alpha\beta}&=&\mathrm{diag}\left[\rho_D, -p^x_{D}, -p^y_D, - p^z_D\right]\\[2mm]
\nonumber
&=&\mathrm{diag}\left[1, -\omega^x_D, -\omega^y_D, -\omega^z_D\right]\rho_D\\[2mm]
&=&\mathrm{diag}\left[1,-\omega_D,-(\omega_D+\alpha),-(\omega_D+\alpha)\right]\rho_D\,,
\end{eqnarray}
respectively, where $p_D$ is the dark energy
pressure and $\omega_D=p_D/\rho_D$ the related
Equation of State (EoS) parameter. 
The deviation from isotropy is parametrized
by setting $\omega^x_D= \omega_D$ and introducing the deviations
along $y$ and $z$ axes by the (time-dependent) skewness parameter $\alpha$. Clearly, the standard isotropic framework is recovered
in the $\alpha\rightarrow0$ limit. 

Let us now consider a homogeneous and anisotropic KS
Universe of metric
\be
\label{SBmetric}
ds^2\,=\,dt^2-A^2(t)\,dr^2-B^2(t)\,(d\theta^2+\sin^2\theta\, d\phi^2)\,,
\ee
where $A$ and $B$ are the (time-dependent)
metric potentials. In this framework, the field equations~\eqref{F1}  become
\begin{eqnarray}
\label{1}
2\frac{\ddot B}{B}+\frac{\dot B^2}{B}+\frac{1}{B^2}-\frac{w}{2}\phi^n\dot\phi^2&=&-\omega_D\rho_D\,,\\[2mm]
\label{2}
\frac{\ddot A}{A}+\frac{\ddot B}{B}+\frac{\dot A \dot B}{A B}-\frac{w}{2}\phi^n\dot\phi^2&=&-(\omega_D+\alpha)\rho_D\,,\\[2mm]
\label{3}
2\frac{\dot A \dot B}{A B}+\frac{\dot B^2}{B^2}+\frac{1}{B^2}+\frac{w}{2}\phi^n\dot\phi^2&=&\rho_m+\rho_D\,,\\[2mm]
\label{4}
\ddot\phi+\left(\frac{\dot A}{A}+2\frac{\dot B}{B}\right)\dot\phi+\frac{n}{2}\frac{\dot\phi^2}{\phi}&=&0\,,
\end{eqnarray}
where the dot denotes ordinary derivative with respect 
to the cosmic time $t$. 

Similarly, the continuity equation~\eqref{cont} can be cast
as
\begin{eqnarray}
\nonumber
&&\dot \rho_m+\left(\frac{\dot A}{A}+2\frac{\dot B}{B}\right)\rho_m+\dot\rho_D+\left(\frac{\dot A}{A}+2\frac{\dot B}{B}\right)\left(1+\omega_D\right)\rho_D\\[2mm]
&&+2\frac{\dot B}{B}\alpha\rho_D=0\,.
\label{nce}
\end{eqnarray}

In order to solve the system of four equations~\eqref{1}-\eqref{4}
in the seven unknowns $A,B,\rho_m,\rho_D,\omega_D$, $\alpha$ and $\phi$, we need to impose some extra conditions. 
Following~\cite{Collins}, we require the metric potentials
to be related by
\be
A\,=\,B^k\,,
\ee
with $k\neq1$ being a positive constant. 
Furthermore, we set~\cite{Skewness1,Skewness2}
\be
\label{skew}
\alpha\,=\,\frac{\alpha_0\left(k-1\right)\dot B B-1}{B^2\rho_D}\,,
\ee
where $\alpha_0$ is an arbitrary constant. 

In so doing, the metric potentials take the form
\begin{eqnarray}
\label{A}
A&=&\left[\frac{\alpha_1\left(k+2\right)}{\alpha_0}e^{\alpha_0t}+\alpha_2(k+2)\right]^{\frac{k}{k+2}}\,,\\[2mm]
B&=&\left[\frac{\alpha_1\left(k+2\right)}{\alpha_0}e^{\alpha_0t}+\alpha_2(k+2)\right]^{\frac{1}{k+2}}\,,
\label{B}
\end{eqnarray}
where $\alpha_1$ and $\alpha_2$ are integration constants. 
Hence, Eq.~\eqref{SBmetric} with the substitution of Eqs.~\eqref{A} and~\eqref{B} describes the geometry of KS Universe
in SBT. 

Finally, we observe that the Hubble parameter
for our model is given by~\cite{Santhi}
\be
\label{HP}
H\,=\,\left(\frac{\dot A}{A}+2\,\frac{\dot B}{B}\right)\,
\ee
where we have absorbed an overall $1/3$ 
in the redefinition of $A$ and $B$.

\subsection{Barrow Holographic Dark Energy}
HDE in its most common formulation avails of 
the Hubble horizon as an IR cutoff and Bekenstein-Hawking
area law for the horizon DoF of Universe. 
However,  its failures to reproduce the whole cosmic
evolution have motivated tentative changes over the years.
Some proposals have been put forward by considering either 
different IR cutoffs~\cite{Wang:2016och, Wang:2016lxa} or
modified horizon entropies~\cite{Tsallis1,Tsallis2,Tsallis3,Tsallis4,Kana1,Kana2,Kana3,Bar1,Bar2,Bar3,Bar4,Bar5,Bar6,Bar7}. Among the latter models,
HDE based on Barrow entropy~\cite{B2020} has 
been attracting great attention in the last years~\cite{Bar1,Bar2,Bar3,Bar4,Bar5,Bar6,Bar7}.

Inspired by the Covid-19 virus structure,
Barrow has proposed that quantum gravity
effects might affect black hole horizon structure, 
introducing intricate, fractal features~\cite{B2020}. 
In turn, this would generalize black hole entropy formula to
\be
\label{BE}
S\sim{A}^{1+\Delta/2}\,, 
\ee
where $0\le\Delta\le1$, with $\Delta=1$ ($\Delta=0$)
corresponding to the maximal (vanishing) deviation from 
the standard entropy-area law. Notice
that observational constraints on $\Delta$ have been derived
in~\cite{Bar6,Anagnostopoulos:2020ctz,Barrow:2020kug,LucianoInf}. 

Based on Eq.~\eqref{BE} and exploiting the deep
connection between gravity and thermodynamics, 
Barrow's paradigm has recently been extended to Cosmology.
Specifically, as argued in~\cite{Bar1} the definition of HDE density in Barrow's picture appears as
\be
\label{rhoD}
\rho_D\,=\,CL^{\Delta-2}\,,
\ee
where $C$ is an unknown parameter with dimensions $[L]^{-2-\Delta}$.
By setting the Hubble horizon as IR cutoff, we then get
\be
\label{rhode}
\rho_D\,=\,CH^{2-\Delta}\,.
\ee 
where $H$ is given by Eq.~\eqref{HP}. From this relation, 
we easily get
\be
\label{dotH}
\dot\rho_D\,=\,C\left(2-\Delta\right)H^{1-\Delta}\dot H\,.
\ee

Equations~\eqref{A},~\eqref{B} and~\eqref{rhode} provide
the necessary tools for our next reconstruction of BHDE in SBT.

\section{BHDE reconstruction in Saez-Ballester Theory}
\label{reco}
Let us now describe the evolution of a KS Universe 
with anisotropic BHDE and dark matter. We analyze separately the
cases where: \emph{i}) there is no energy exchange between
the cosmos sectors and \emph{ii}) a suitable 
interaction is assumed to exist.

\subsection{Non-interacting model}
\label{NIM}
As a first step, we observe that in this model the energy
conservation equations for dark matter and BHDE
can be decoupled to give
\begin{eqnarray}
\label{cont1}
&&\dot\rho_m+3H\rho_m\,=\,0\,,\\[2mm]
&&\dot\rho_D+3H(1+\omega_D)\rho_D+2\alpha\rho_D\frac{\dot B}{B}\,=\,0\,.
\label{cont2}
\end{eqnarray}
respectively. 
Also, from Eq.~\eqref{4} we obtain
\begin{eqnarray}
\nonumber
\hspace{-3mm}\phi^{\frac{n+2}{2}}&=&\left(\frac{n+2}{2}\right)\phi_0\\[2mm]
&&\hspace{-12mm}\times\,\int\left[\frac{\alpha_1\left(k+2\right)}{\alpha_0}\,e^{\alpha_0t}+\alpha_2\left(k+2\right)
\right]^{-3} dt\,+\,\phi_1\,,
\label{phin2}
\end{eqnarray}
where $\phi_0$ and $\phi_1$ are integration constants. 

By plugging Eqs.~\eqref{B},~\eqref{rhode} and~\eqref{phin2}
into~\eqref{3}, we are led to
\begin{eqnarray}
\nonumber
\rho_m&=&\frac{\left(2k+1\right)\alpha_0^2\,\alpha_1^2\,e^{2\alpha_0t}}{\left(k+2\right)^2\left(\alpha_1 e^{\alpha_0 t}+\alpha_0\alpha_2\right)^2}\\[2mm]
\nonumber
&&+\,\left[\frac{\left(k+2\right)\left(\alpha_1\,e^{\alpha_0t}+\alpha_0\alpha_2\right)}{\alpha_0}\right]^{-\frac{2}{k+2}}\\[2mm]
\nonumber
&&-\,C\left(\frac{\alpha_0\alpha_1}{\alpha_1+\alpha_0\alpha_2\,e^{-\alpha_0t}}\right)^{2-\Delta}\\[2mm]
&&+\,\frac{w\,\alpha_0^6\,\phi_0^2}{2\left(k+2\right)^6\left(\alpha_1\,e^{\alpha_0 t}+\alpha_0\alpha_2\right)^6}\,.
\end{eqnarray}

\begin{figure}[t]
\begin{center}
\includegraphics[width=8.5cm]{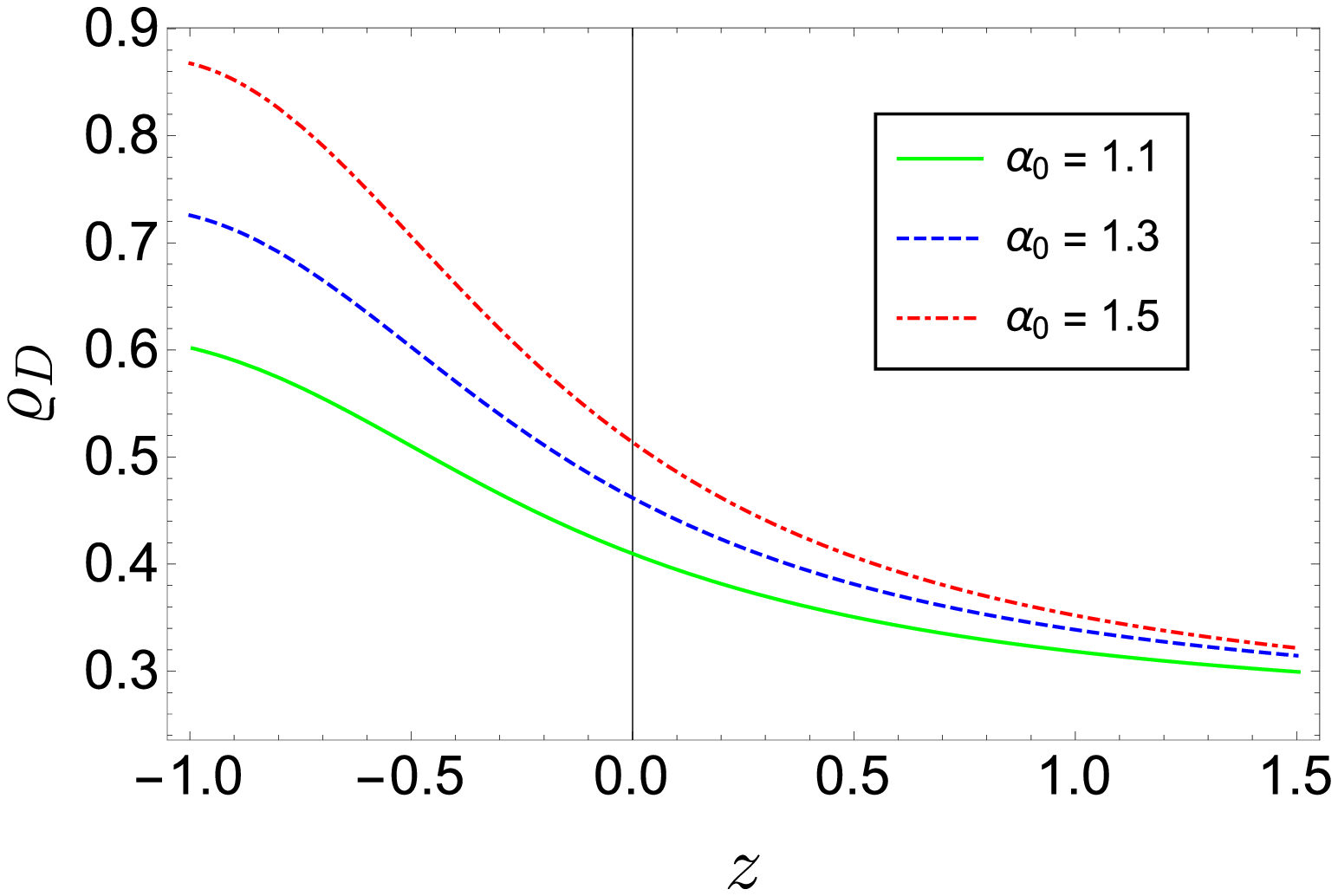}
\end{center}
\begin{center}
\includegraphics[width=8.5cm]{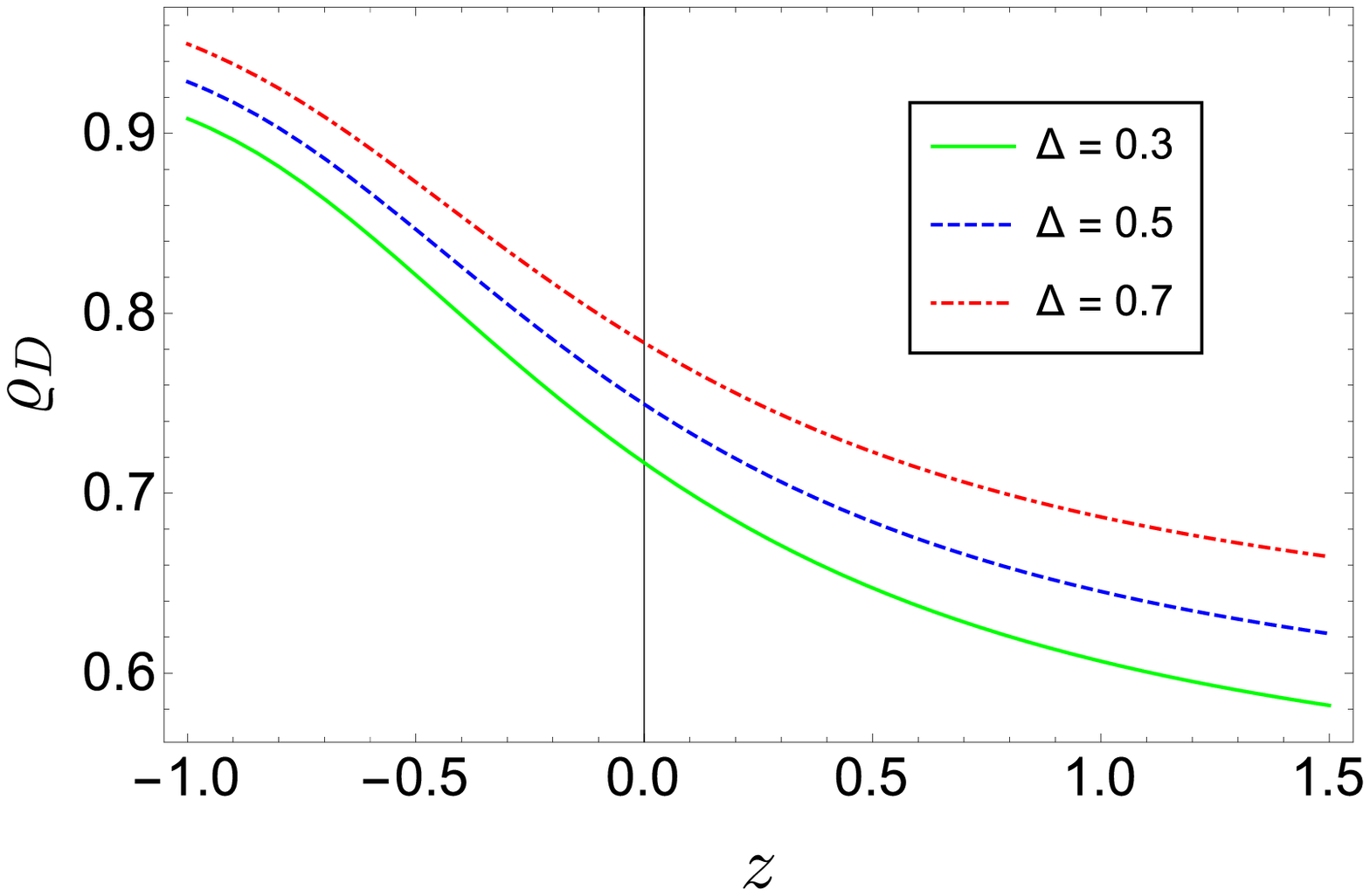}
\caption{Evolution of $\rho_D$ versus $z$ for different values
of $\alpha_0$ (upper panel) and $\Delta$ (lower panel) in non-interacting model. We have set $C=0.1$, $k=1.1$, $\alpha_1=0.9$, $\alpha_2=0.52$, $w=10$, $\phi_0=5$ and $\Delta=0.5$ in the upper panel, while $\alpha_0=1.1$ in the lower one (online colors). }
\label{Fig1}
\end{center}
\end{figure}

In order to understand the role of skewness 
and Barrow parameters in the evolution of the Universe, 
we now analyze the dynamics of model parameters
for various values of $\alpha_0$ and $\Delta$.
The evolution of $\rho_D$ versus the redshift $z$ 
is plotted in Fig.~\ref{Fig1}. 
One can see that BHDE increases as $z$ decreases 
and comes to dominate the energy budget of the Universe in the far future. Also, $\rho_D$
increases with increasing $\alpha_0$ (see the upper panel of Fig.~\ref{Fig1}), while it is only slightly affected by variation of $\Delta$ (see the lower panel of Fig.~\ref{Fig1}). 

\begin{figure}[t]
\begin{center}
\includegraphics[width=8.5cm]{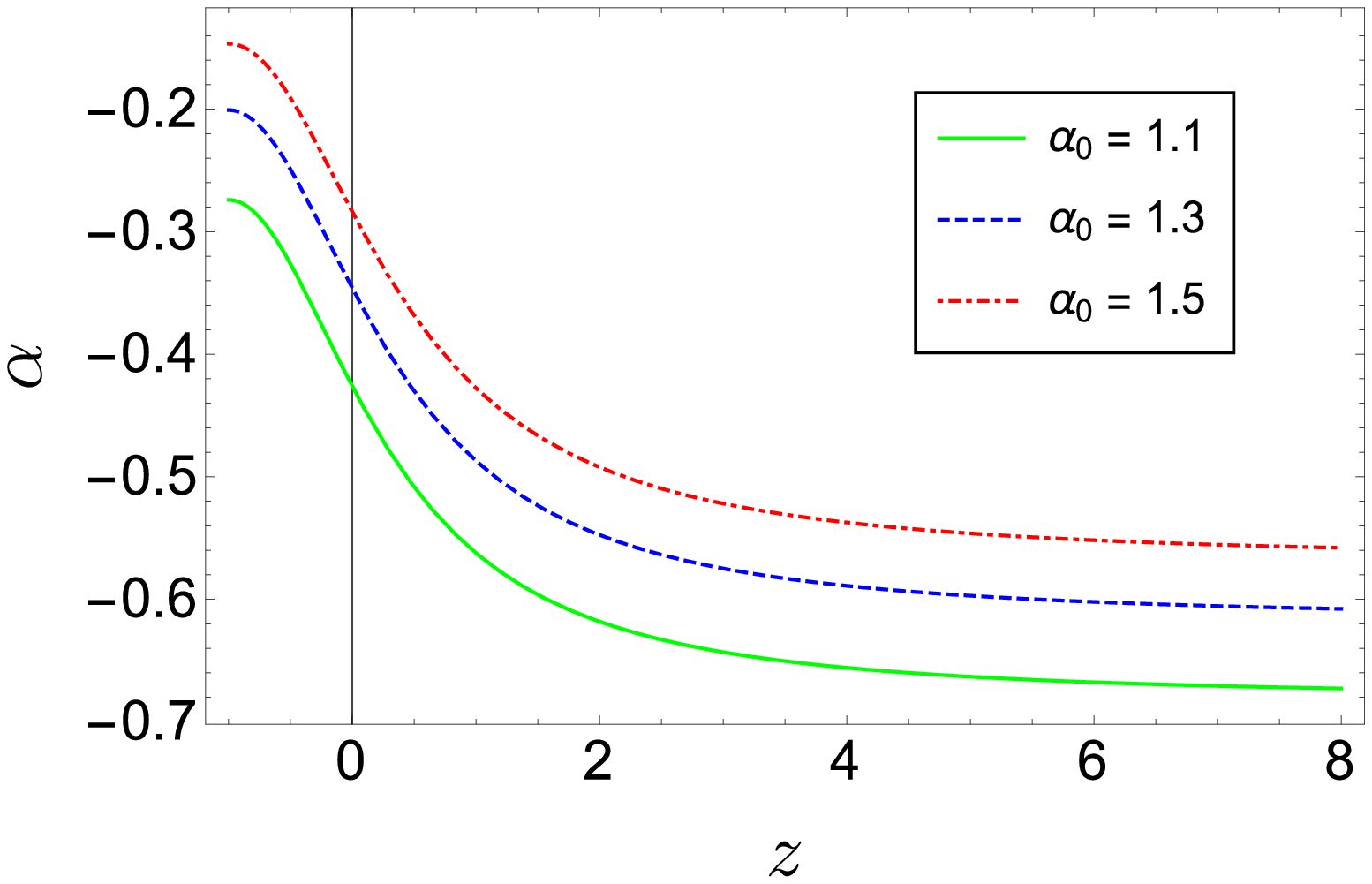}
\end{center}
\begin{center}
\includegraphics[width=8.5cm]{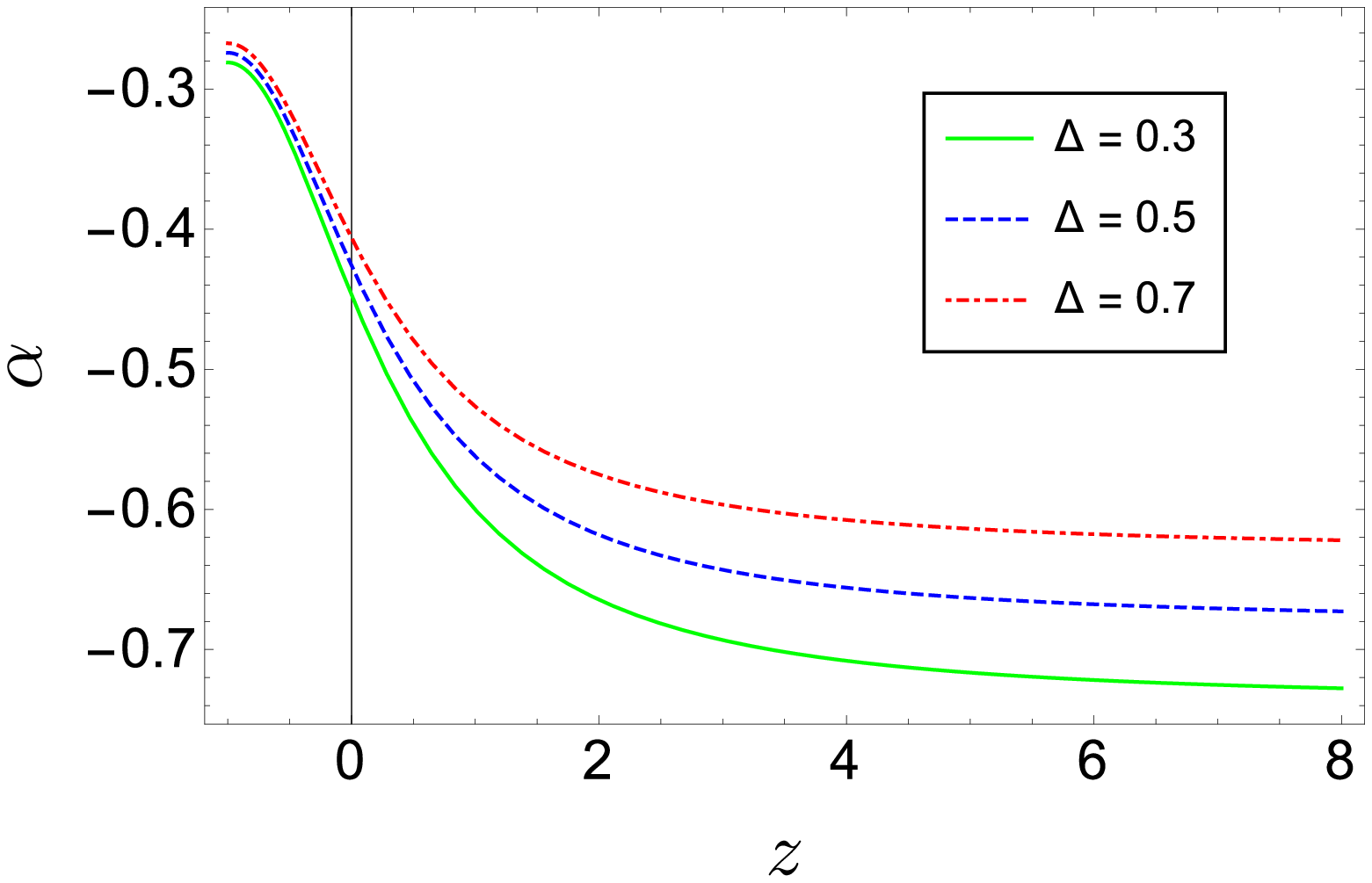}
\caption{Evolution of skewness parameter versus $z$ for different values
of $\alpha_0$ (upper panel) and $\Delta$ (lower panel) in non-interacting model. For all model parameters, we have set the same values as in Fig.~\ref{Fig1} (online colors).}
\label{Fig3}
\end{center}
\end{figure}

In Fig.~\ref{Fig3} we depict the evolution of 
the skewness parameter~\eqref{skew}. We observe
that it is negative and approaches constant values
in the far future for selected values of $\alpha_0$ (upper panel) and $\Delta$ (lower panel). The same asymptotic behavior
is exhibited in~\cite{Raju} for the case of a Kantowski-Sachs cosmological model with anisotropic dark energy fluid and massive scalar field. 

Now, from Eqs.~\eqref{B},~\eqref{rhode} and~\eqref{cont2}
we can derive the expression of the EoS parameter
of BHDE as
\begin{eqnarray}
\nonumber
\omega_D&=&-1\,-\frac{2\,e^{3\alpha_0 t}\,\alpha_0^5\,\alpha_1^{\Delta-2}
\left(\frac{\alpha_0}{\alpha_1+e^{-\alpha_0t}\,\alpha_0\alpha_2}\right)^{-\Delta}}
{3C\left(k+2\right)^2\left(\alpha_1\,e^{\alpha_0 t}+\alpha_0\alpha_2\right)^5}\\[2mm]
\nonumber
&&\hspace{-10mm}\times\,
\left\{\left(k-1\right)e^{\alpha_0t}\,\alpha_0\alpha_1-
\left[\frac{\left(k+2\right)\left(\alpha_1\,e^{\alpha_0t}+\alpha_0\alpha_2\right)}{\alpha_0}
\right]^{\frac{k}{2+k}}
\right\}\\[2mm]
&&\hspace{-10mm}+\,C\,e^{\alpha_0t}\,\alpha_0\,\alpha_1\,\alpha_2
\left(\frac{\alpha_0\alpha_1}{\alpha_1+e^{-\alpha_0t}\alpha_0\alpha_2}\right)^{1-\Delta}\left(\Delta-2\right).
\label{OmDNI}
\end{eqnarray}

\begin{figure}[t]
\begin{center}
\includegraphics[width=8.5cm]{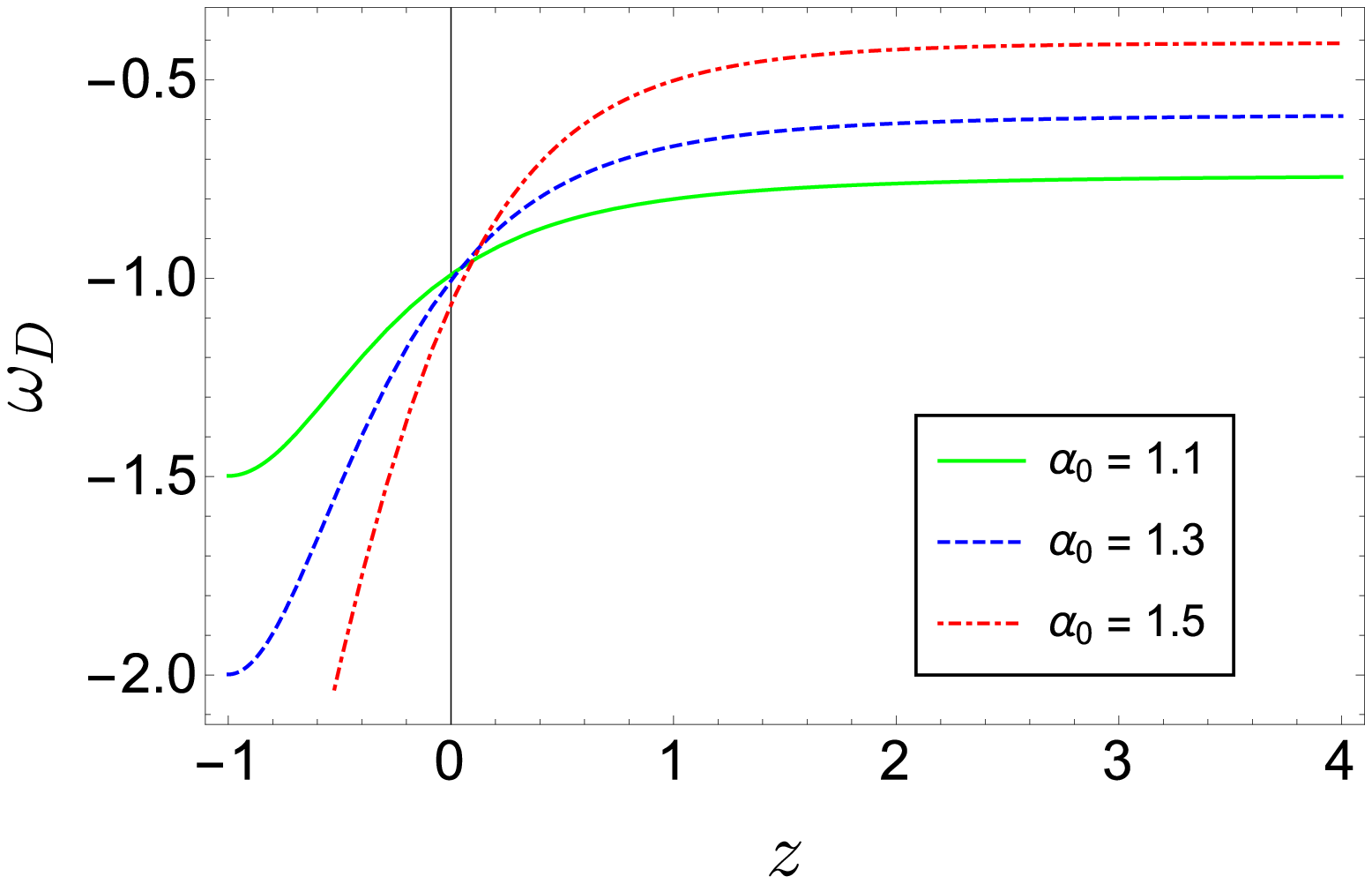}
\end{center}
\begin{center}
\includegraphics[width=8.5cm]{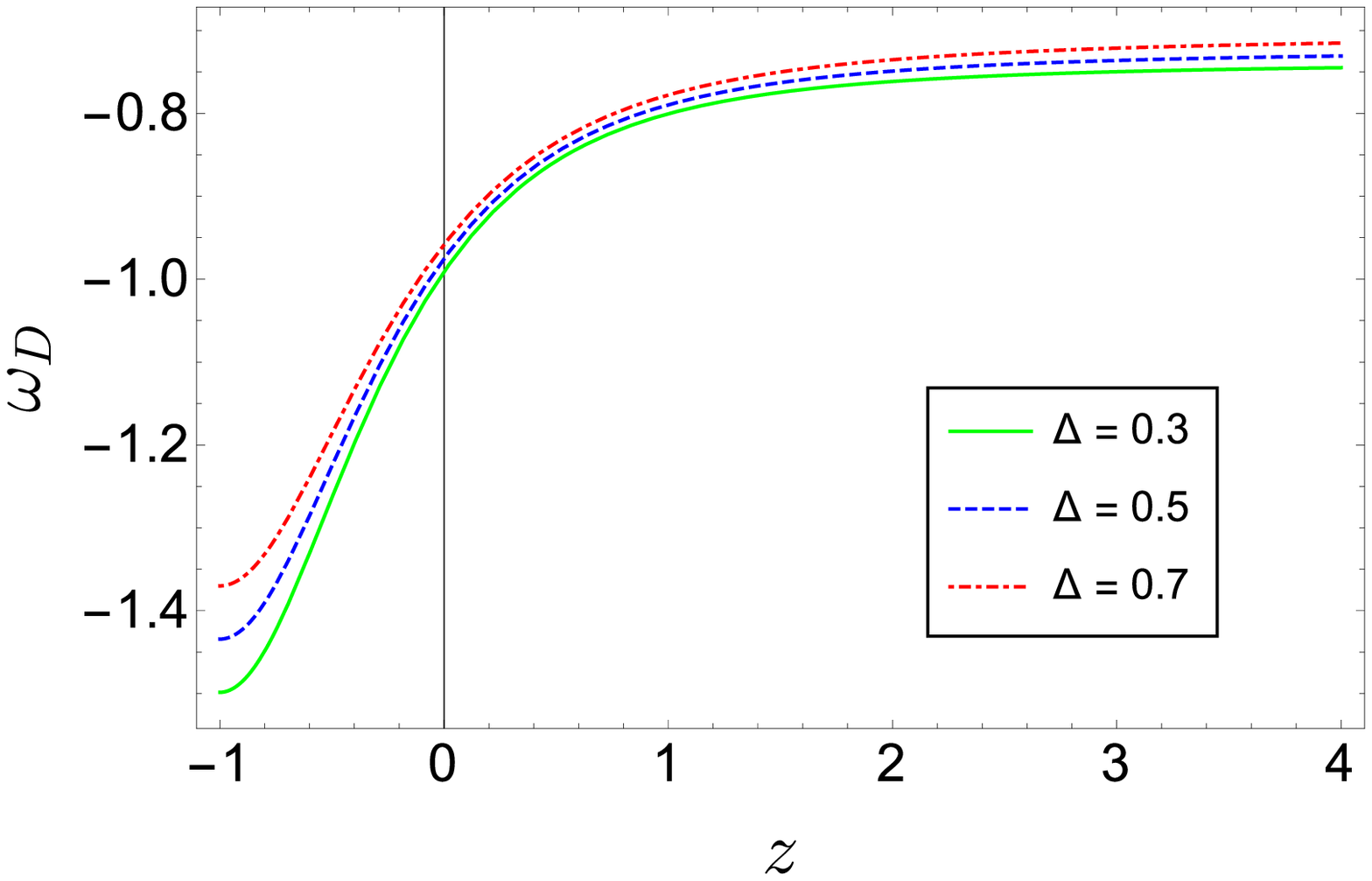}
\caption{Evolution of EoS parameter versus $z$ for different values
of $\alpha_0$ (upper panel) and $\Delta$ (lower panel) in non-interacting mode. For all model parameters, we have set the same values as in Fig.~\ref{Fig1} (online colors).}
\label{Fig4}
\end{center}
\end{figure}

This is plotted in Fig.~\ref{Fig4}: from the upper panel we see that
BHDE evolves from quintessence ($-1<\omega_D<-1/3$) at late time
to approximately cosmological constant ($\omega_D=-1$) at present and phantom
($\omega_D<-1$) in the far future. In this regard, it is worth 
noting that largely negative values of $\omega_D$ 
indicate that the Universe might either end up
with a big-rip or remain in the same
current accelerating status. 

By comparison with results of~\cite{Santhi}, 
we infer that the obtained behavior is peculiar to BHDE
model in KS Universe. In fact, Tsallis HDE always
lies in a quintessence-like regime and approaches
cosmological constant in the far future. On the other hand, 
the same evolution is exhibited in the context of BHDE in Brans-Dicke Cosmology with a linear interaction~\cite{Bar7} and Bianchi-type I BHDE
in teleparallel gravity~\cite{TeleBar}. Furthermore, 
quantitative analysis
gives us $\omega_{D_0}\in [-1.07, -0.99]$ for the current value of the EoS parameter and the considered values of $\alpha_0$. This is in good agreement with the recent constraint $\omega_0\in[-1.38,-0.89]$ obtained from Planck+WP+BAO measurements~\cite{Planck}. A qualitatively similar transition from quintessence to cosmological constant
and phantom is displayed in the lower panel of Fig.~\ref{Fig4}
for fixed $\alpha_0$ and varying $\Delta$. In this case
we find $\omega_{D_0}\in[-0.99,-0.95]$, which is still consistent
with observations~\cite{Planck}. 

\begin{figure}[t]
\begin{center}
\includegraphics[width=8.5cm]{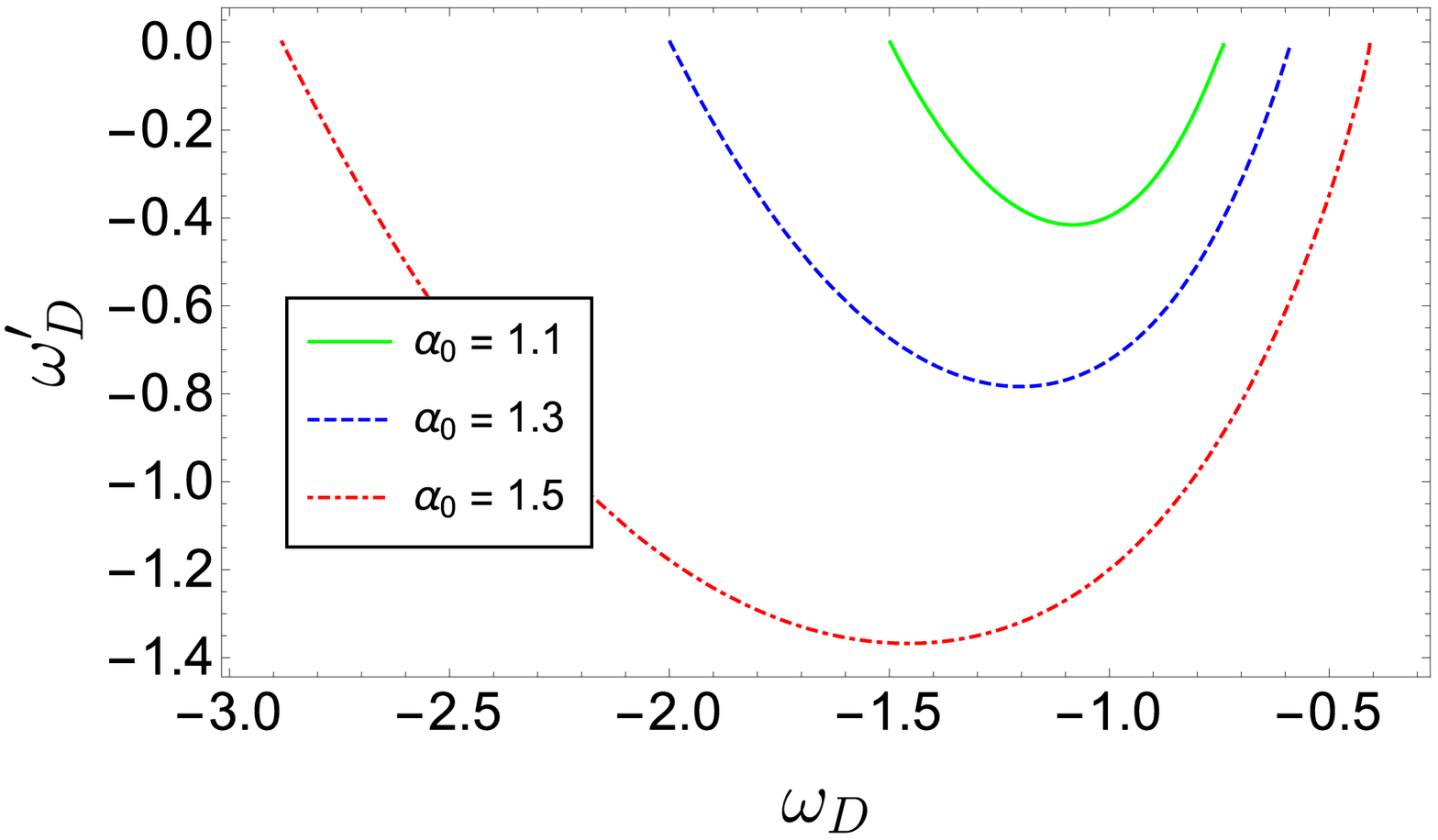}
\end{center}
\begin{center}
\includegraphics[width=8.5cm]{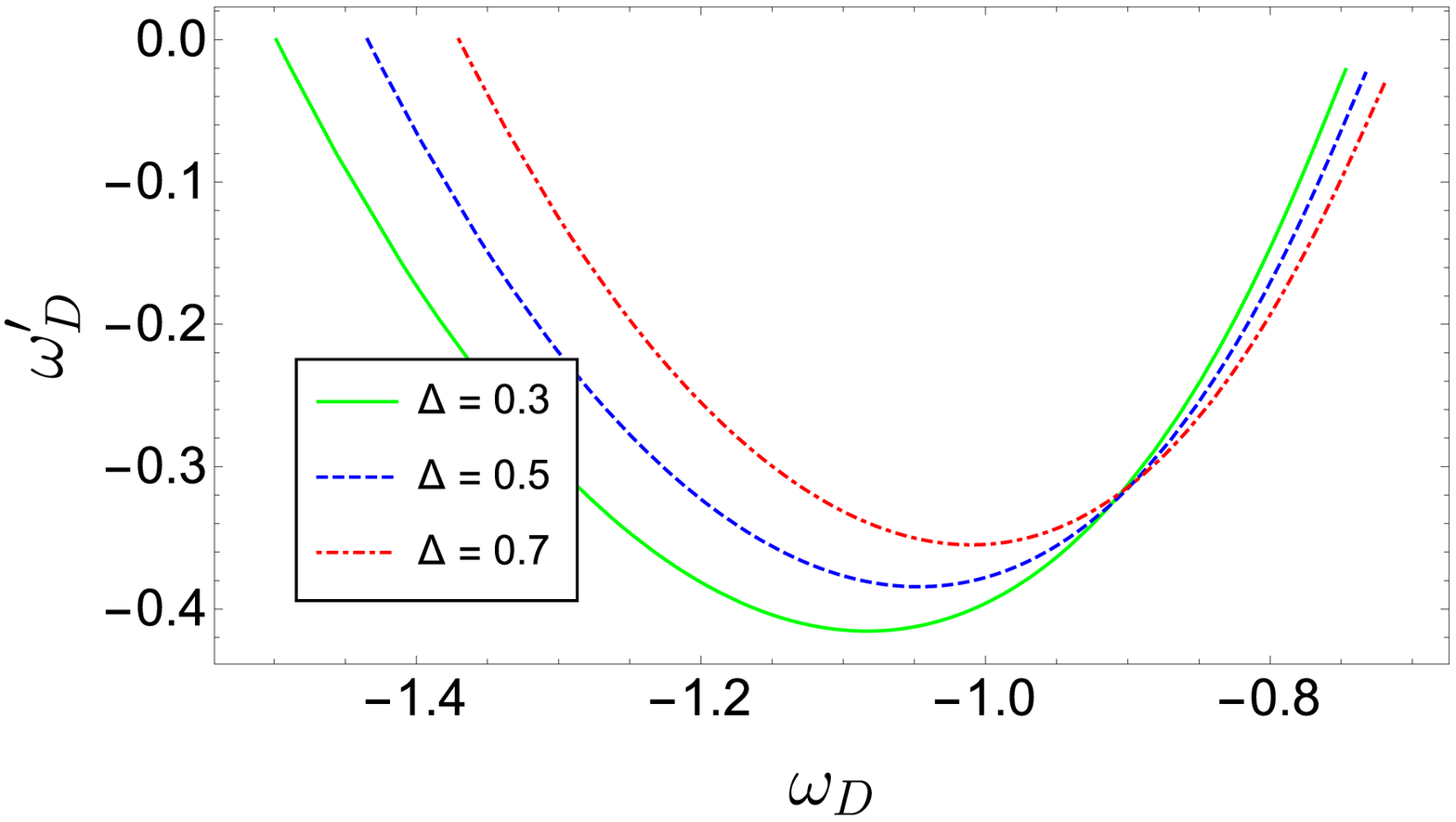}
\caption{Evolution of $\omega_D-\omega'_D$ trajectories 
for different values
of $\alpha_0$ (upper panel) and $\Delta$ (lower panel) in non-interacting model. For all model parameters, we have set the same values as in Fig.~\ref{Fig1} (online colors).}
\label{Fig5}
\end{center}
\end{figure}

Let us now investigate trajectories of $\omega_D-\omega_D'$ plane. Here the overhead prime denotes derivative respect to the logarithm of the scale factor $a$, which is as usual related to the Hubble parameter  by $H=\dot a/a$. $\omega_D-\omega_D'$ plane has been introduced
by Caldwell and Linder~\cite{CL} and represents
a useful tool to distinguish among dark energy models. 
Firstly, it has been applied to quintessence model, 
which gives two different regions in this plane, i.e.
the \emph{thawing} ($\omega_{D}<0,\omega_{D}'>0$)
and {\emph {freezing}} ($\omega_{D}<0,\omega_{D}'<0$) domains.
Subsequently, it has been generalized and extended to other
dynamical dark energy models, see for instance~\cite{Sche,Chiba,Guo,Sharif}. 
Cosmic trajectories of $\omega_D-\omega_D'$ plane
for the present framework are plotted in Fig.~\ref{Fig5}.
We observe that our model predicts freezing region for BHDE,  
which is consistent with the current behavior of the Universe, 
since freezing regime is associated to a more accelerating
era of cosmic expansion respect to thawing domain.  
The same result is exhibited in~\cite{TeleBar} for the case of BHDE in teleparallel gravity~\cite{TeleBar}, while
the opposite scenario occurs in Tsallis HDE in KS Universe~\cite{Santhi}.

Another quantity that should be taken into account
to establish whether a dark energy model is 
phenomenologically consistent is the deceleration parameter
\be
\label{qeq}
q\,=\,-\frac{\ddot a}{aH^2}\,=\,-1-\frac{\dot H}{H^2}\,.
\ee

\begin{figure}[t]
\begin{center}
\includegraphics[width=8.5cm]{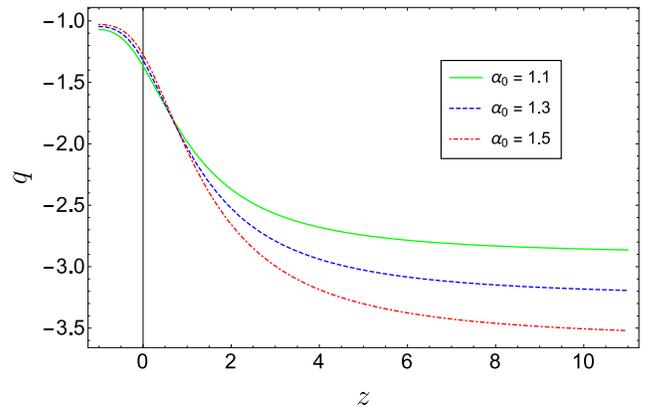}
\caption{Evolution of deceleration parameter for different values
of $\alpha_0$ in non-interacting model. For all model parameters, we have set the same values as in Fig.~\ref{Fig1}  (online colors).}
\label{Fig6}
\end{center}
\end{figure}

From this equation, we infer that positive values of $q$
correspond to a decelerated expansion of the Universe
($\ddot a<0$), while negative values characterize
the accelerated regime ($\ddot a>0$). The behavior of $q$
is displayed in Fig.~\ref{Fig6}, showing that
our model  correctly reproduces the
current accelerating phase of the cosmos.
We emphasize that this is an 
advantage of BHDE scenario
over standard HDE, which by contrast fails to 
explain the present accelerated expansion.
Quantitatively speaking, for the selected values
of $\alpha_0$ we find $q_0\in[-1.35,-1.23]$ for the current deceleration
parameter. Although it slightly deviates from the standard
$\Lambda$CDM model value $q_0=-0.55$~\cite{Planck}, it 
overlaps with the estimation $q_0\in[-1.37,-0.79]$ recently
obtained in~\cite{Camerana} via local supernovae measurements.

\begin{figure}[t]
\begin{center}
\includegraphics[width=8.5cm]{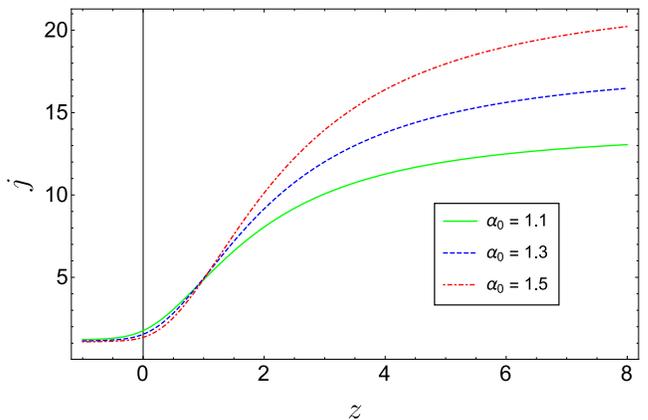}
\caption{Evolution of jerk parameter for different values
of $\alpha_0$ in non-interacting model. For all model parameters, we have set the same values as in Fig.~\ref{Fig1} (online colors).}
\label{Fig7}
\end{center}
\end{figure}

In Fig.~\ref{Fig7} we present the evolution of the
jerk parameter, which is a dimensionless third 
derivative of the scale factor
respect to the cosmic time, i.e.~\cite{Blandford:2004ah}
\begin{eqnarray}
\nonumber
j&=&\frac{1}{aH^3}\frac{d^3a}{dt^3}\,=\,q\left(2q+1\right)+\left(1+z\right)\frac{dq}{dz}\\[2mm]
\nonumber
&&\hspace{-5mm}=\,1\,+\frac{9e^{-\frac{3\alpha_0}{1+(1+z)^2}}\alpha_0\alpha_2\left\{
\left[2+z\left(2+z\right)\right]^2-2\left(1+z\right)^2\alpha_0
\right\}}{\left[2+z\left(2+z\right)\right]^2\alpha_1}\\[2mm]
&&\hspace{-5mm}+\,\frac{18e^{-\frac{6\alpha_0}{1+\left(1+z\right)^2}}\alpha_0^2\,\alpha_2^2}{\alpha_1^2}\,.
\end{eqnarray}

We point out that the this parameter allows us to quantify deviations
from $\Lambda$CDM model, which is indeed characterized
by $j=1$. From Fig.~\ref{Fig7} we can see
that our model departs from $\Lambda$CDM at early times, 
while consistency is achieved in the far future. Also, we have
$j_0\in[1.34,1.83]$ for the selected values of $\alpha_0$.

\begin{figure}[t]
\begin{center}
\includegraphics[width=8.5cm]{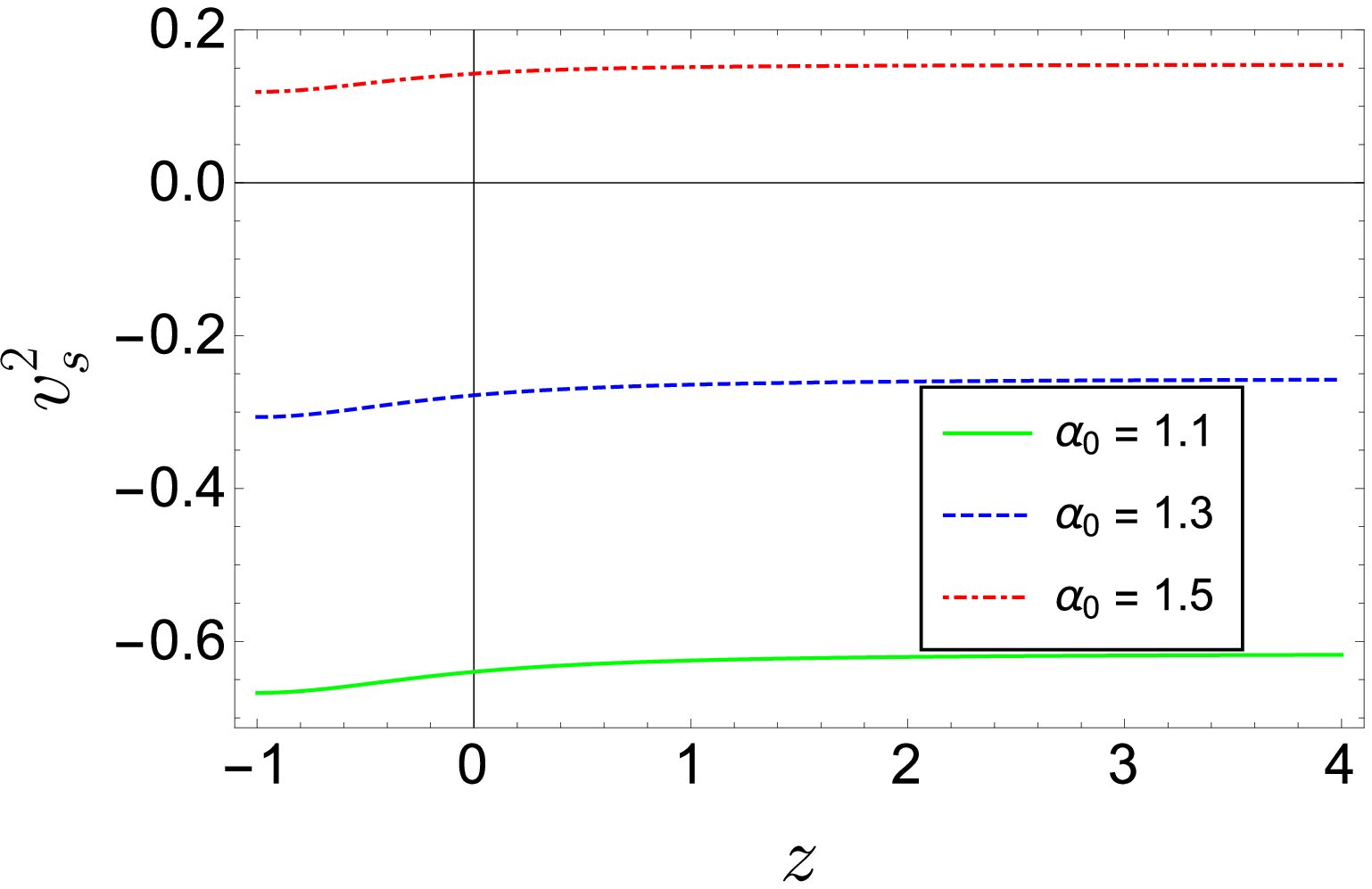}
\end{center}
\begin{center}
\includegraphics[width=8.5cm]{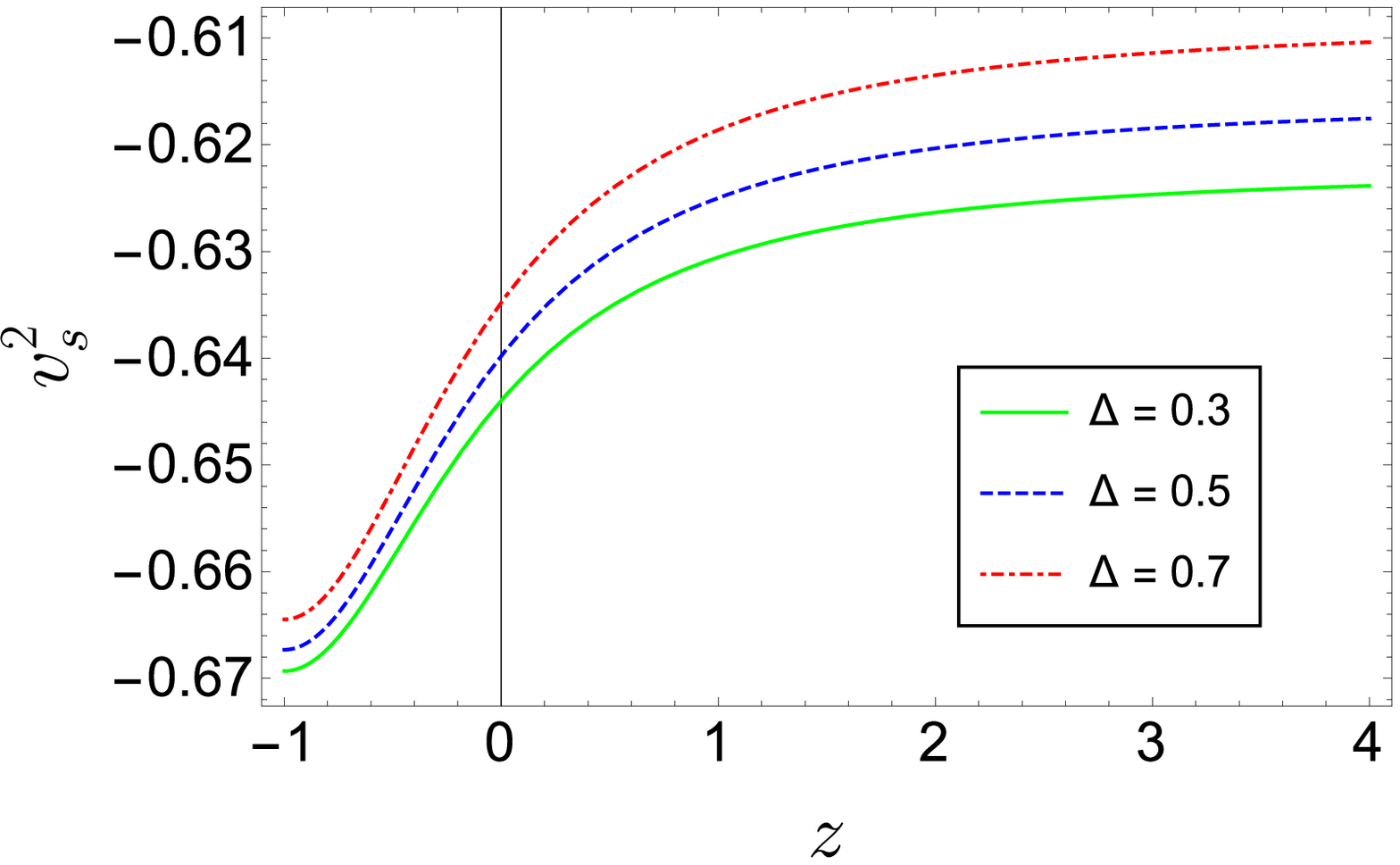}
\caption{Evolution of squared
speed of sound for different values
of $\alpha_0$ (upper panel) and $\Delta$ (lower panel) in non-interacting model. For all model parameters, we have set the same values as in Fig.~\ref{Fig1} (online colors).}
\label{Fig8}
\end{center}
\end{figure}

\begin{figure}[t]
\begin{center}
\includegraphics[width=8.5cm]{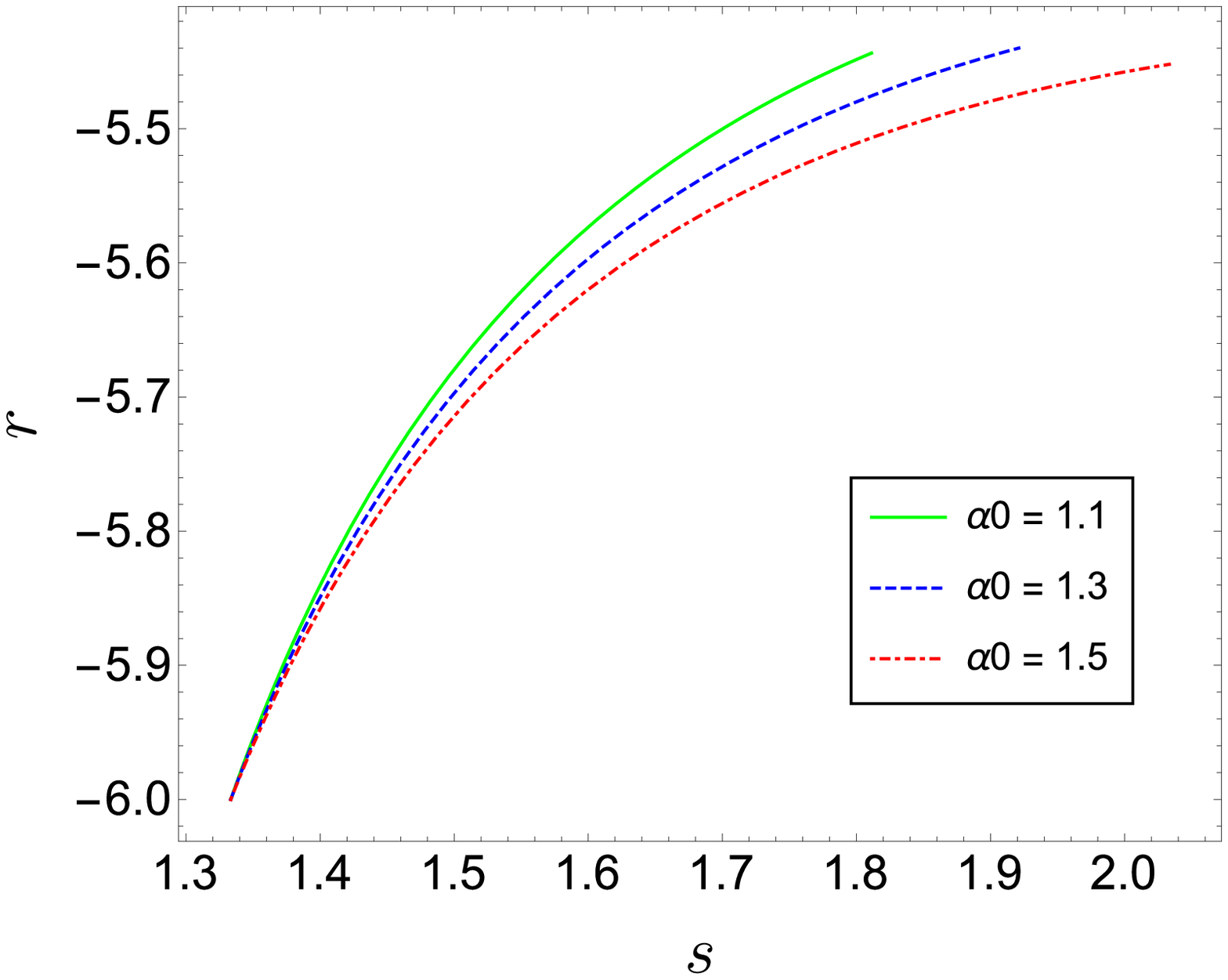}
\caption{Evolution of $(r,s)$ plane trajectories for different values
of $\alpha_0$. For all model parameters, we have set the same values as in Fig.~\ref{Fig1} (online colors).}
\label{Fig8bis}
\end{center}
\end{figure}

In order to study the classical stability of our model against small perturbations, let us now evaluate the squared
speed of sound
\be
\label{vsq}
v_s^2\,=\,\frac{\dot p_{D}}{\dot\rho_{D}}\,.
\ee

Notice that, in order for a given dark energy model to be stable, 
we must have $v_s^2>0$. Indeed, for a density perturbation,
positive values of $v_s^2$ correspond to
a regular propagation mode. 
On the other hand, for $v_s^2<0$ one has 
that the perturbation equation becomes
an irregular wave equation, giving rise to 
an escalating mode~\cite{Pert1}. 
In this setting, the pressure turns out to decrease
when the density perturbation increases, 
thus favoring the development 
of an instability.

For the present BHDE model, we obtain the non-trivial 
expression
\begin{widetext}
\begin{eqnarray}
\nonumber
v_s^2&=&-1+C\alpha_0\,\alpha_1\,\alpha_2\left(\Delta-2\right)e^{\alpha_0t}H^{1-\Delta}\,+\,
\frac{2\,\alpha_1^{2\Delta-5}}{3C}\bigg\{1-\alpha_0\alpha_1\left(k-1\right)e^{\alpha_0t}\left[\frac{\alpha_1(k+2)}{\alpha_0}\,e^{\alpha_0t}+\alpha_2\left(k+2\right)\right]^{-\frac{k}{2+k}}\bigg\}\\[2mm]
\nonumber
&&\hspace{-9mm}\times\,\left[\frac{\alpha_1(k+2)}{\alpha_0}\,e^{\alpha_0t}+\alpha_2\left(k+2\right)\right]^{-\frac{4+k}{2+k}}\,H^{3-\Delta}\,+\,
C\alpha_0\,\alpha_1\,\alpha_2\,\left(\Delta-2\right)e^{\alpha_0t}H^{-\left(1+\Delta\right)}\left[\left(1-\Delta\right)\dot H+\alpha_0\,H\right]\frac{H^{1-\Delta}}{(2-\Delta)\dot H}\\[2mm]
\nonumber
&&\hspace{-9mm}+\,\frac{2\,\alpha_1^{2\Delta-5}H^{4-\Delta}}{3C\left(2-\Delta\right)\dot H}\left[\frac{\alpha_1(k+2)}{\alpha_0}\,e^{\alpha_0t}+\alpha_2\left(k+2\right)\right]^{-\frac{4+k}{2+k}}
\Bigg\{\bigg\{1-\alpha_0\,\alpha_1\left(k-1\right)e^{\alpha_0t}
\left[\frac{\alpha_1(k+2)}{\alpha_0}\,e^{\alpha_0t}+\alpha_2\left(k+2\right)\right]^{-\frac{k}{2+k}}\bigg\}\\[2mm]
\nonumber
&&\hspace{-9mm}\times\,\bigg\{\left(3-\Delta\right)\frac{\dot H}{H}-\left(\frac{4+k}{2+k}\right)
\left[\frac{\alpha_1(k+2)}{\alpha_0}\,e^{\alpha_0t}+\alpha_2\left(k+2\right)\right]^{-1}\bigg\}+\alpha_0^2\,\alpha_1\left(1-k\right)e^{\alpha_0t}
\bigg\{
\left[\frac{\alpha_1(k+2)}{\alpha_0}\,e^{\alpha_0t}+\alpha_2\left(k+2\right)\right]^{-\frac{k}{2+k}}\\[2mm]
&&\hspace{-9mm}-\,\frac{\alpha_1\,k}{\alpha_0}\,e^{\alpha_0t}
\left[\frac{\alpha_1(k+2)}{\alpha_0}\,e^{\alpha_0t}+\alpha_2\left(k+2\right)\right]^{-\frac{k+1}{2+k}}\bigg\}
\Bigg\}\,.
\end{eqnarray}
\end{widetext}

This is plotted in Fig.~\ref{Fig8} for different values
of $\alpha_0$ (upper panel) and $\Delta$ (lower panel). 
From the upper panel, we see that the model is classically
stable throughout the whole evolution for higher skewness (red curve), 
while it exhibits the opposite behavior as skewness decreases
(green and blue curves), no matter the value of Barrow parameter (see lower panel). By contrast, SBT-based reconstruction
of Tsallis HDE, as well as non-interacting BHDE in Brans-Dicke Cosmology are always unstable~\cite{Santhi}. This is a further
advantage of our reconstruction.

Before moving onto the study of the interacting model, 
we focus on the statefinder diagnosis of BHDE. 
The statefinder parameters $r$ and $s$ were first
introduced in~\cite{rs} to differentiate
among the plethora of dark energy models. 
In the definition
of these parameters, derivatives of the scale factor exceed
the second order. In particular, we have
\begin{eqnarray}
r&\equiv&\frac{\dddot a}{aH^3}\,, \\[2mm]
s&\equiv& \frac{r-1}{3(q-\frac{1}{2})}\,.
\end{eqnarray}

We remind that the evolutionary trajectories of dark energy models in the $(r,s)$ plane can be classified as quintessence if $r<1$ and
$s>0$ or Chaplygin gas if $r>1$ and $s<0$~\cite{class}. For the present
model we obtain~\cite{Santhi}
\begin{eqnarray}
\nonumber
r&=&-\frac{\alpha_0^2\alpha_2\,e^{\alpha_0t}}{\alpha_1^2}
\left[\frac{\alpha_1}{\alpha_0}+\alpha_2^2\left(1+2\,e^{-\alpha_0t}\right)\right]\\[2mm]
&&+\,\frac{3\alpha_0\alpha_2\,e^{-\alpha_0t}}{\alpha_1}-6\,,\\[2mm]
\nonumber
s&=&-\frac{2\alpha_0^2\alpha_2\,e^{-\alpha_0t}}{3\alpha_1\left(2\alpha_0\alpha_2\,e^{-\alpha_0t}-3\alpha_1\right)}\left[\frac{\alpha_1}{\alpha_0}+\alpha_2^2\left(1+2\,e^{-\alpha_0t}\right)\right]\\[2mm]
&&+\frac{2\alpha_0\alpha_1\alpha_2\,e^{-\alpha_0t}\,-\,4\alpha_1}{\left(2\alpha_0\alpha_2\,e^{-\alpha_0t}-3\alpha_1\right)},
\end{eqnarray}

The trajectories of $(r,s)$ plane are plotted in Fig.~\ref{Fig8bis}, 
indicating that BHDE in this framework
gives a correspondence with quintessence model.

\subsection{Interacting model}
\label{SCM}
Let us now examine how the above framework
gets modified when considering a more realistic scenario 
with interacting dark matter and BHDE. In this case
the continuity equation~\eqref{nce} can be split 
into the two relations
\begin{eqnarray}
\label{cont3}
&&\dot\rho_m+3H\rho_m\,=\,Q\,,\\[2mm]
&&\dot\rho_D+3H(1+\omega_D)\rho_D+2\alpha\rho_D\frac{\dot B}{B}\,=\,-Q\,,
\label{cont4}
\end{eqnarray}
with $Q$ being the interaction term. 

While there is no natural guidance from
fundamental physics on the form of $Q$,
phenomenological arguments have led to explore
many possible scenarios over the years~\cite{Q1,Q2,Q3,Q4,ComT0,ComT1,ComT2}.
Following~\cite{ComT0,ComT1,ComT2}, here we assume\footnote{
Notice that $\rho$ in Eq.~\eqref{inter} might be in
principle $\rho_m$, $\rho_D$ or even $\rho_{tot}$. For consistency with~\cite{Santhi}, we set $\rho=\rho_D$. However, the same considerations can be carried out by resorting to the more general interaction $Q = 3 H \left(b_1^2 \rho_m + b_2^2 \rho_D\right)$, with $b_{1,2}$ being dimensionless constant much less than unity and positive, so as to be consistent with the Le Chatelier-Braun principle. While being characterized by one more free parameter, we expect that the new framework could yield similar results. A more detailed analysis of this point is left for future investigation.}
\be
\label{inter}
Q=3\beta Hq\rho_D\,, 
\ee
where $\beta$ is a dimensionless constant, which should
take negative values according to observational measurements~\cite{ComT1}.

\begin{figure}[t]
\begin{center}
\includegraphics[width=8.5cm]{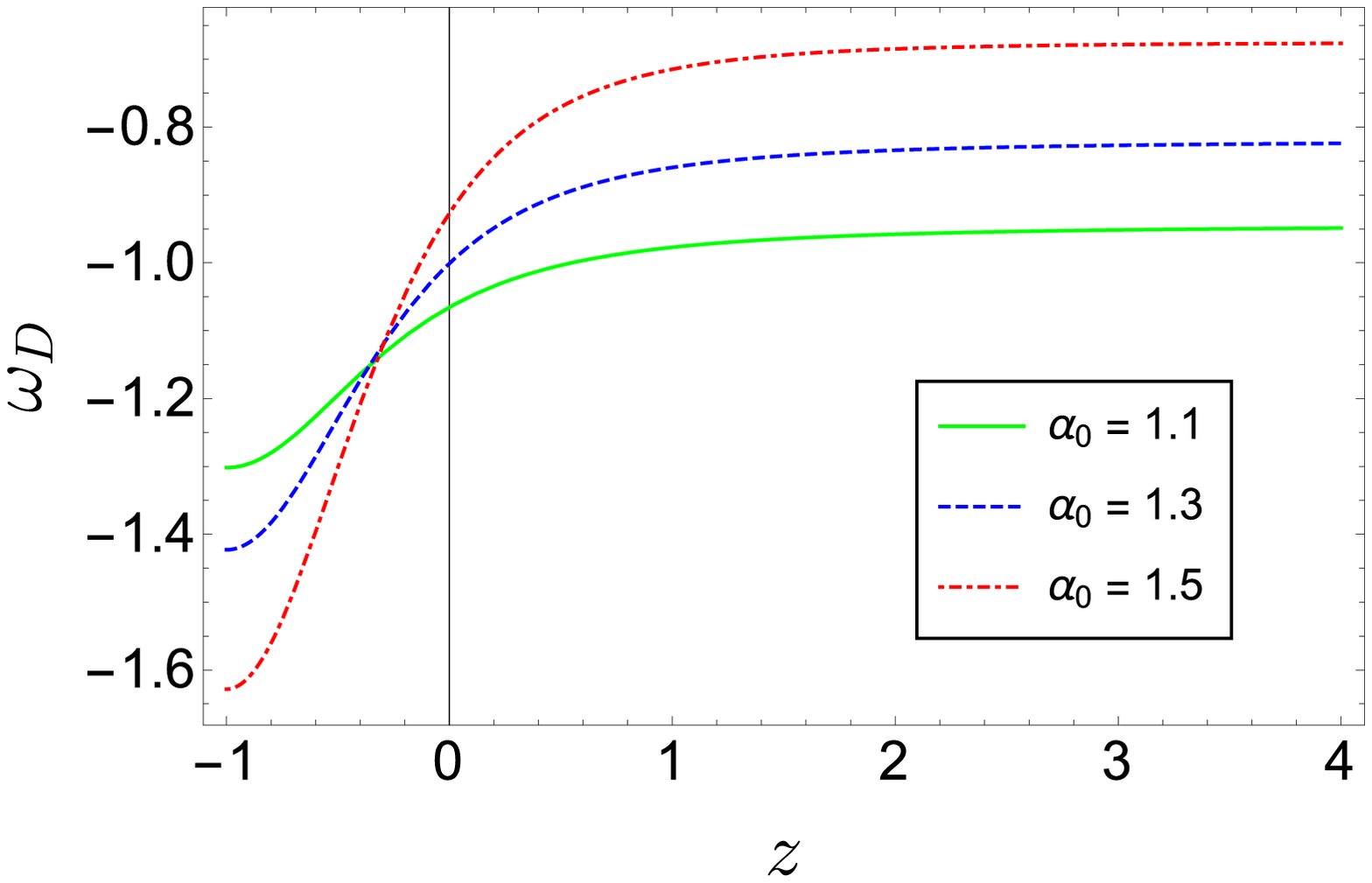}
\end{center}
\begin{center}
\includegraphics[width=8.5cm]{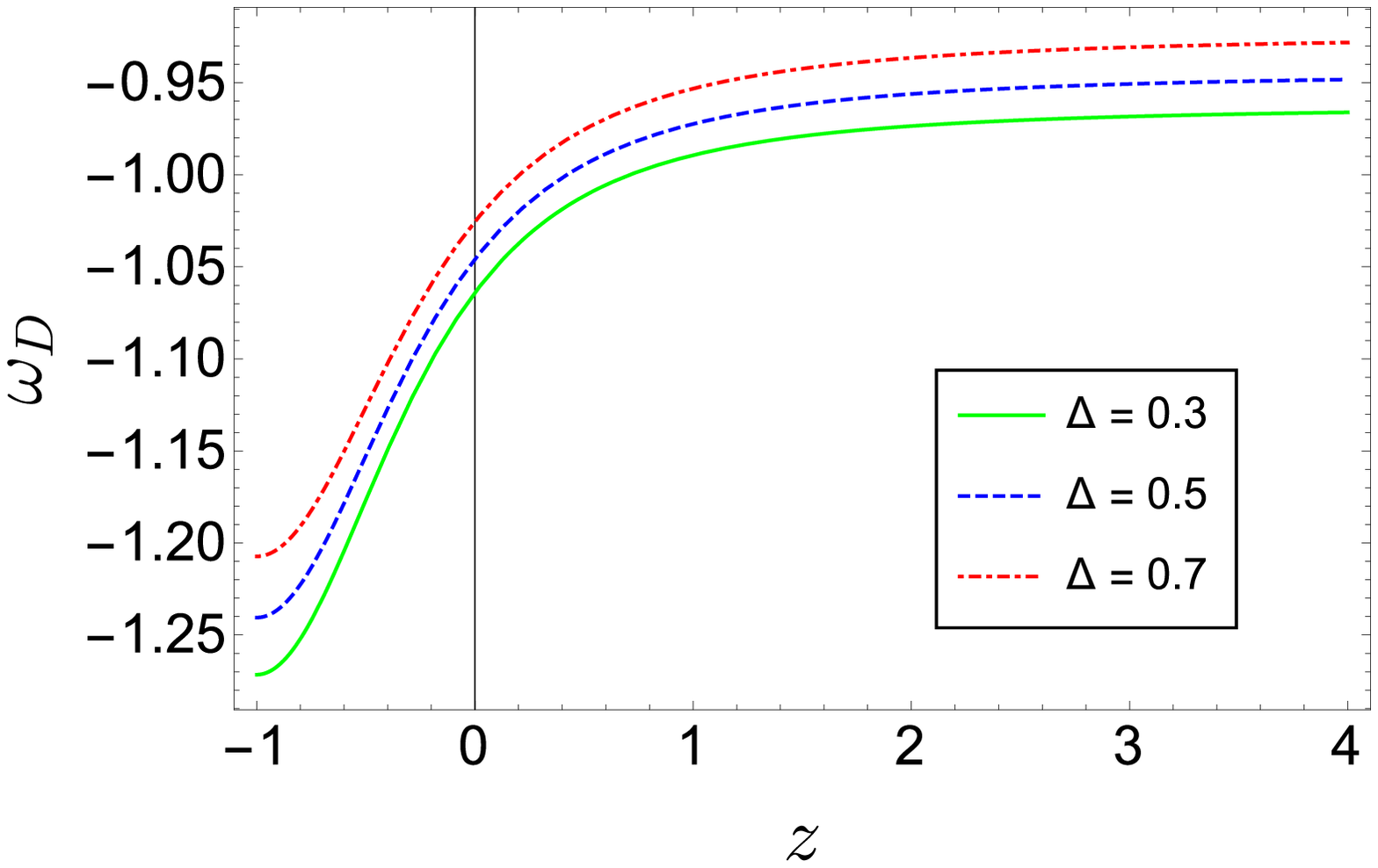}
\end{center}
\begin{center}
\includegraphics[width=8.5cm]{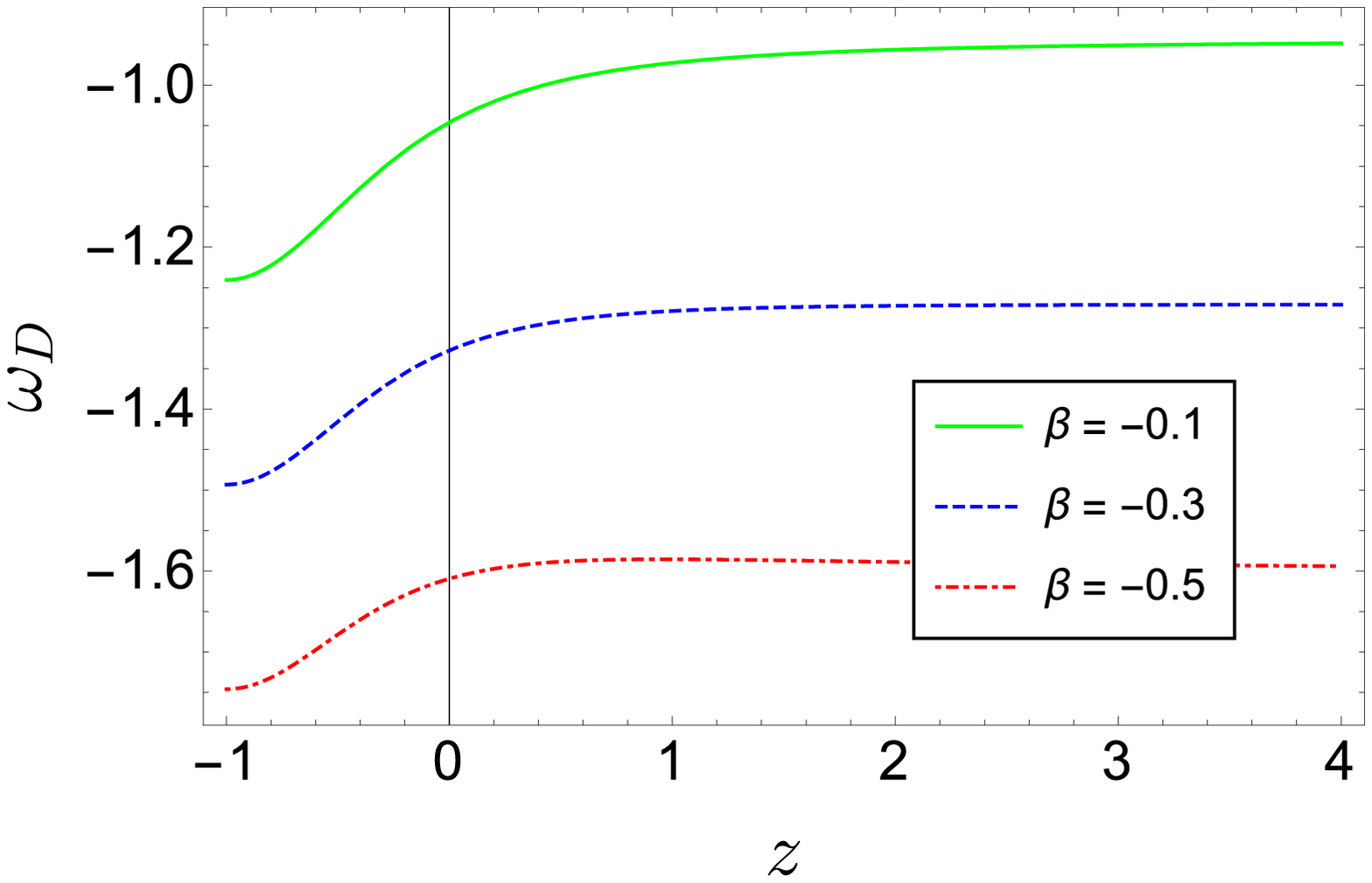}
\caption{Evolution of EoS parameter for different values
of $\alpha_0$ (upper panel), $\Delta$ (middle panel) and $\beta$ (lower panel) in interacting model. For all model parameters, we have set the same values as in Fig.~\ref{Fig1}. For the upper and middle panels, we have considered $\beta=-0.1$ (online colors).}
\label{Fig9}
\end{center}
\end{figure}

Within this framework, the EoS parameter for dark energy
becomes
\begin{eqnarray}
\nonumber
\omega_D&=&-1\,-\frac{2\,e^{3\alpha_0 t}\,\alpha_0^5\,\alpha_1^{\Delta-2}
\left(\frac{\alpha_0}{\alpha_1+e^{-\alpha_0t}\,\alpha_0\alpha_2}\right)^{-\Delta}}
{3C\left(k+2\right)^2\left(\alpha_1\,e^{\alpha_0 t}+\alpha_0\alpha_2\right)^5}\\[2mm]
\nonumber
&&\hspace{-10mm}\times\,
\left\{\left(k-1\right)e^{\alpha_0t}\,\alpha_0\alpha_1-
\left[\frac{\left(k+2\right)\left(\alpha_1\,e^{\alpha_0t}+\alpha_0\alpha_2\right)}{\alpha_0}
\right]^{\frac{k}{2+k}}
\right\}\\[2mm]
\nonumber
&&\hspace{-10mm}+\,C\,e^{\alpha_0t}\,\alpha_0\,\alpha_1\,\alpha_2
\left(\frac{\alpha_0\alpha_1}{\alpha_1+e^{-\alpha_0t}\alpha_0\alpha_2}\right)^{1-\Delta}\left(\Delta-2\right)\,-\,\beta q\,.\\
\end{eqnarray}

\begin{figure}[H]
\begin{center}
\includegraphics[width=7.5cm]{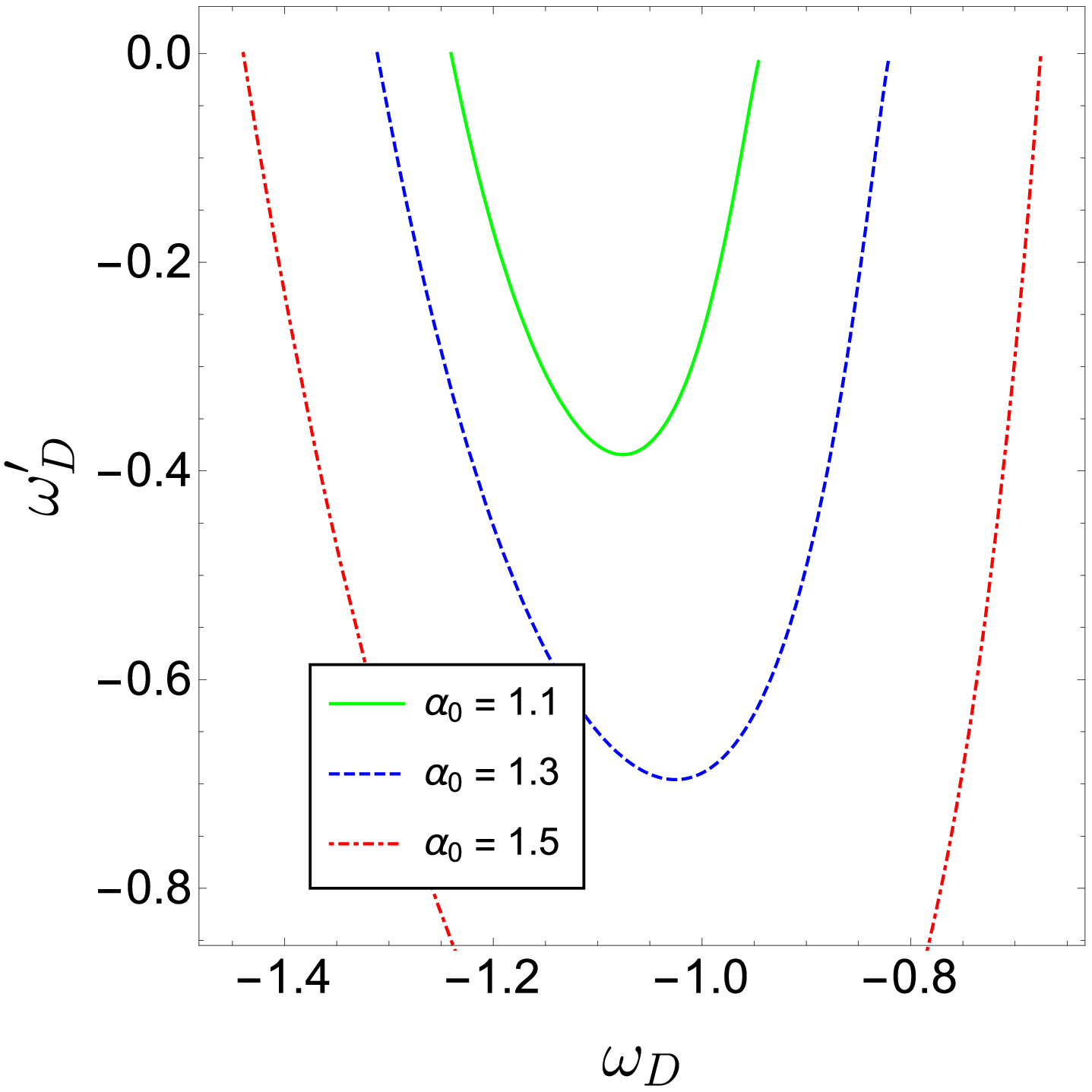}
\end{center}
\begin{center}
\includegraphics[width=8.5cm]{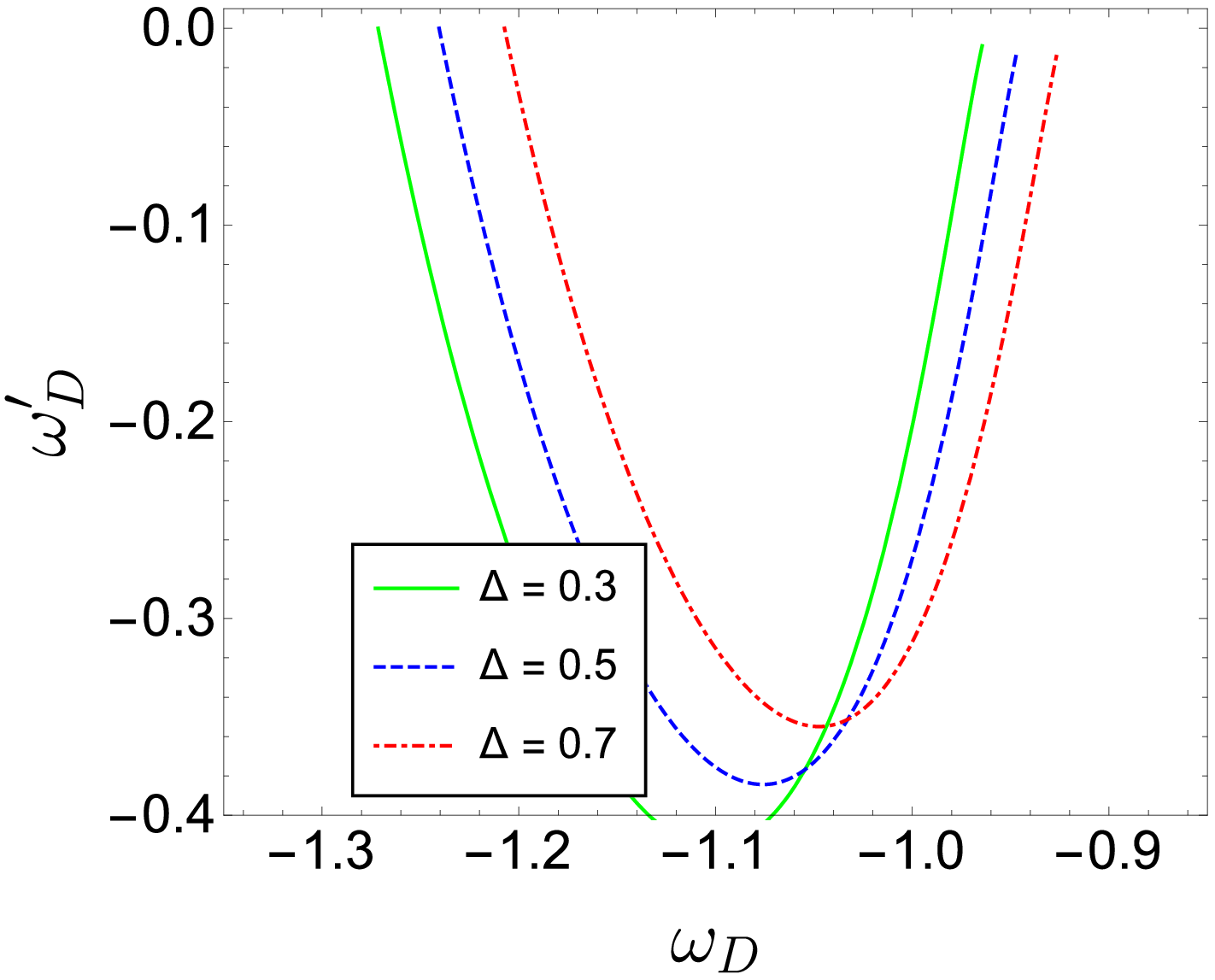}
\end{center}
\begin{center}
\includegraphics[width=8.5cm]{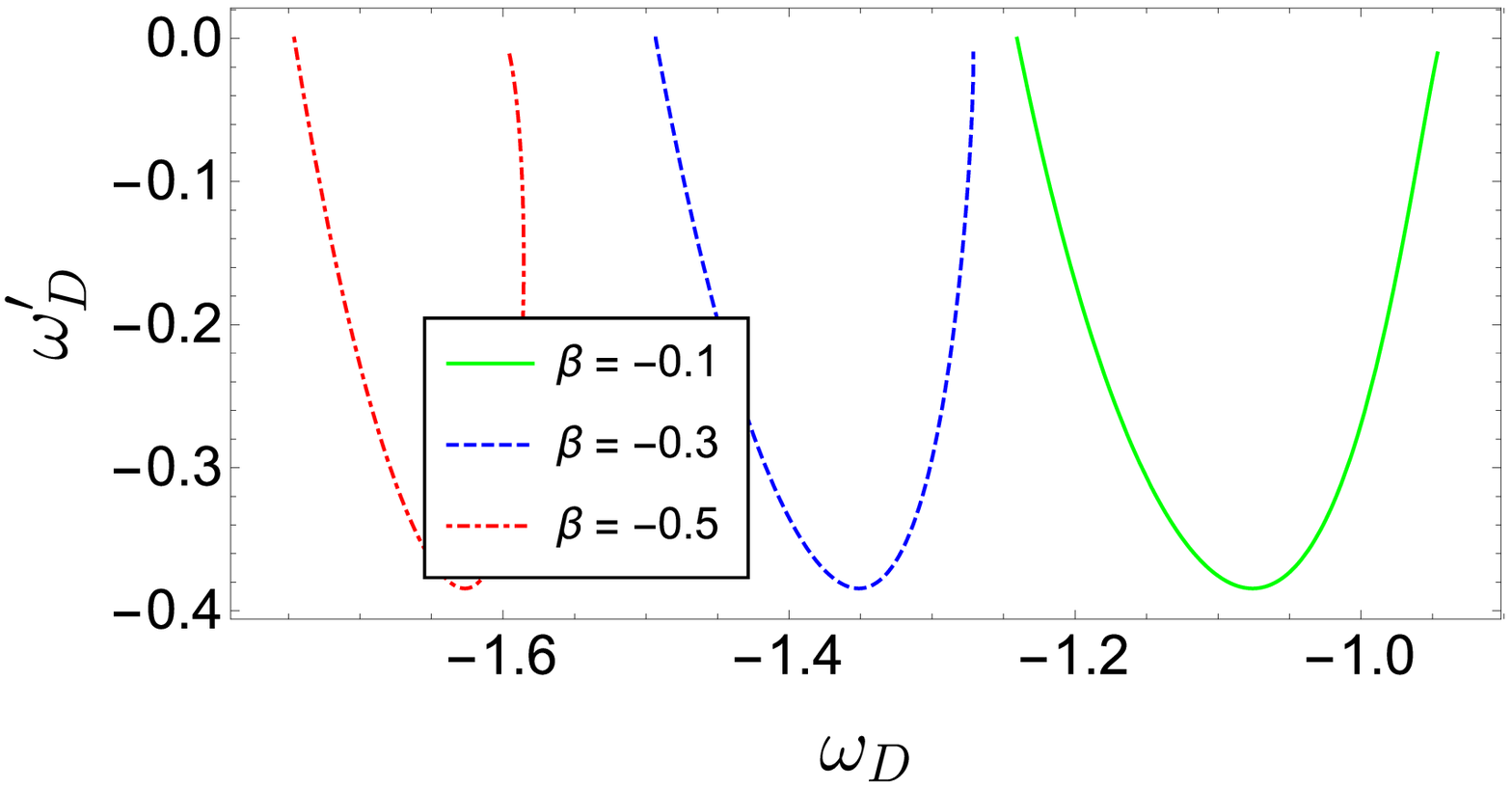}
\caption{
Evolution of $\omega_D-\omega'_D$ trajectories 
for different values
of $\alpha_0$ (upper panel), $\Delta$ (middle panel) and $\beta$ (lower panel) panel in interacting model. For all model parameters, we have set the same values as in Fig.~\ref{Fig1}. For the upper and middle panels, we have considered $\beta=-0.1$ (online colors).}
\label{Fig10}
\end{center}
\end{figure}

For interaction small enough, 
the predicted evolution is qualitatively similar to the
previous model, with the sequence of quintessence-, cosmological
constant- and phantom-like behaviors (see upper and middle panels of Fig.~\ref{Fig9}). Estimation
of the present value of $\omega_D$ now 
gives $\omega_{D_0}\in [-1.06,-0.92]$ for fixed $\Delta=0.5, \beta=-0.1$
and varying $\alpha_0$ (upper panel), and 
$\omega_{D_0}\in [-1.06,-1.02]$ for fixed $\alpha_0=1.1, \beta=-0.1$ and varying $\Delta$ (middle panel). 
Both these ranges are still in good agreement with observations (see the
discussion below Eq.~\eqref{OmDNI}). 
However, by increasing the magnitude of $\beta$, 
we find that BHDE always lies in the phantom regime (see blue
and red curves in the lower panel), yielding
$\omega_{D_0}\in [-1.61,-1.04]$. Therefore, we
infer that large values of $\beta$ are  phenomenologically
disfavored, in line with the result of~\cite{ComT1}.

Figure~\ref{Fig10} displays the trajectories
of $\omega_D-\omega'_D$ phase plane. As for non-interacting model, 
they show that BHDE in SBT lies in the freezing domain.

\begin{figure}{}
\begin{center}
\includegraphics[width=8.5cm]{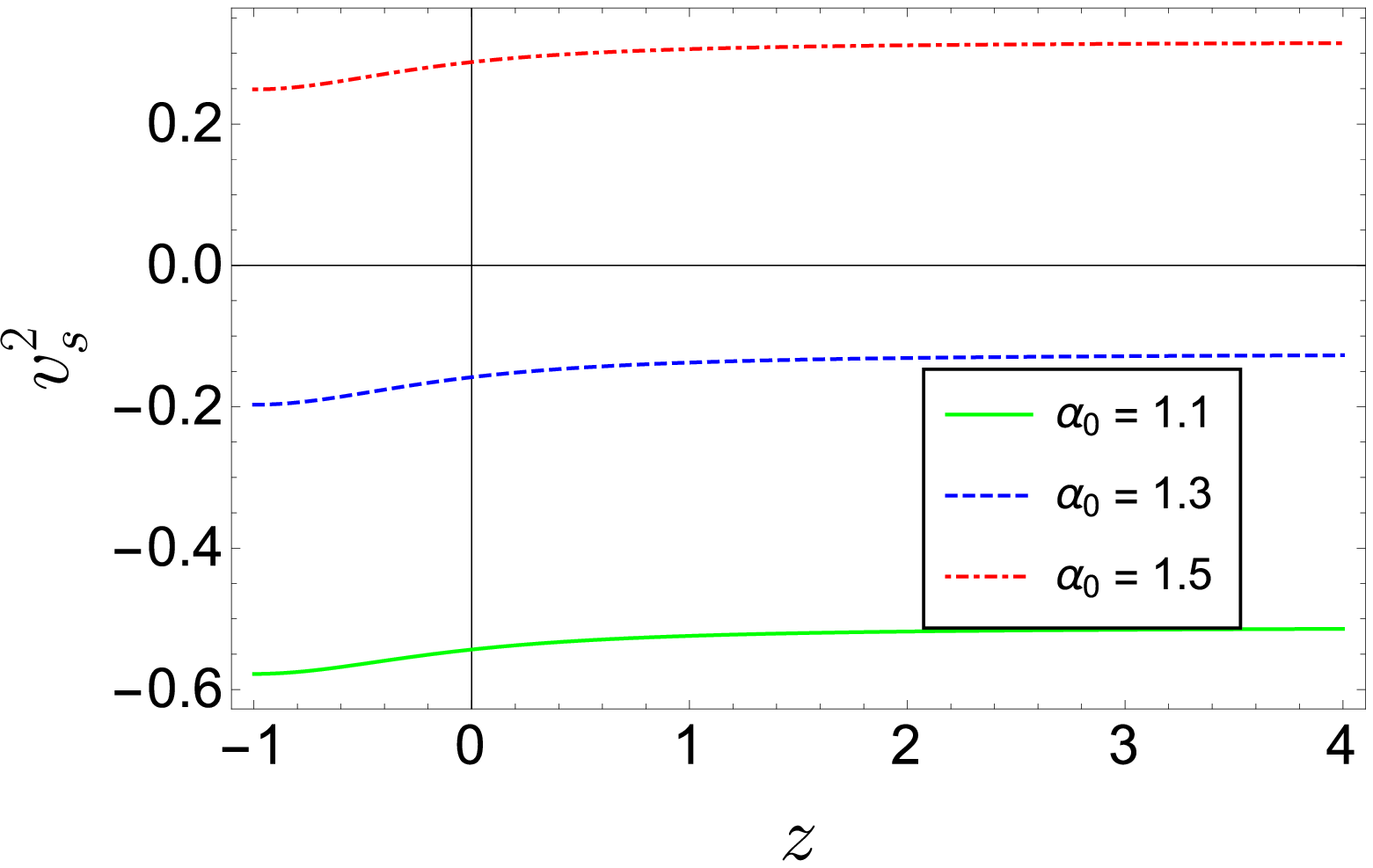}
\end{center}
\begin{center}
\includegraphics[width=8.5cm]{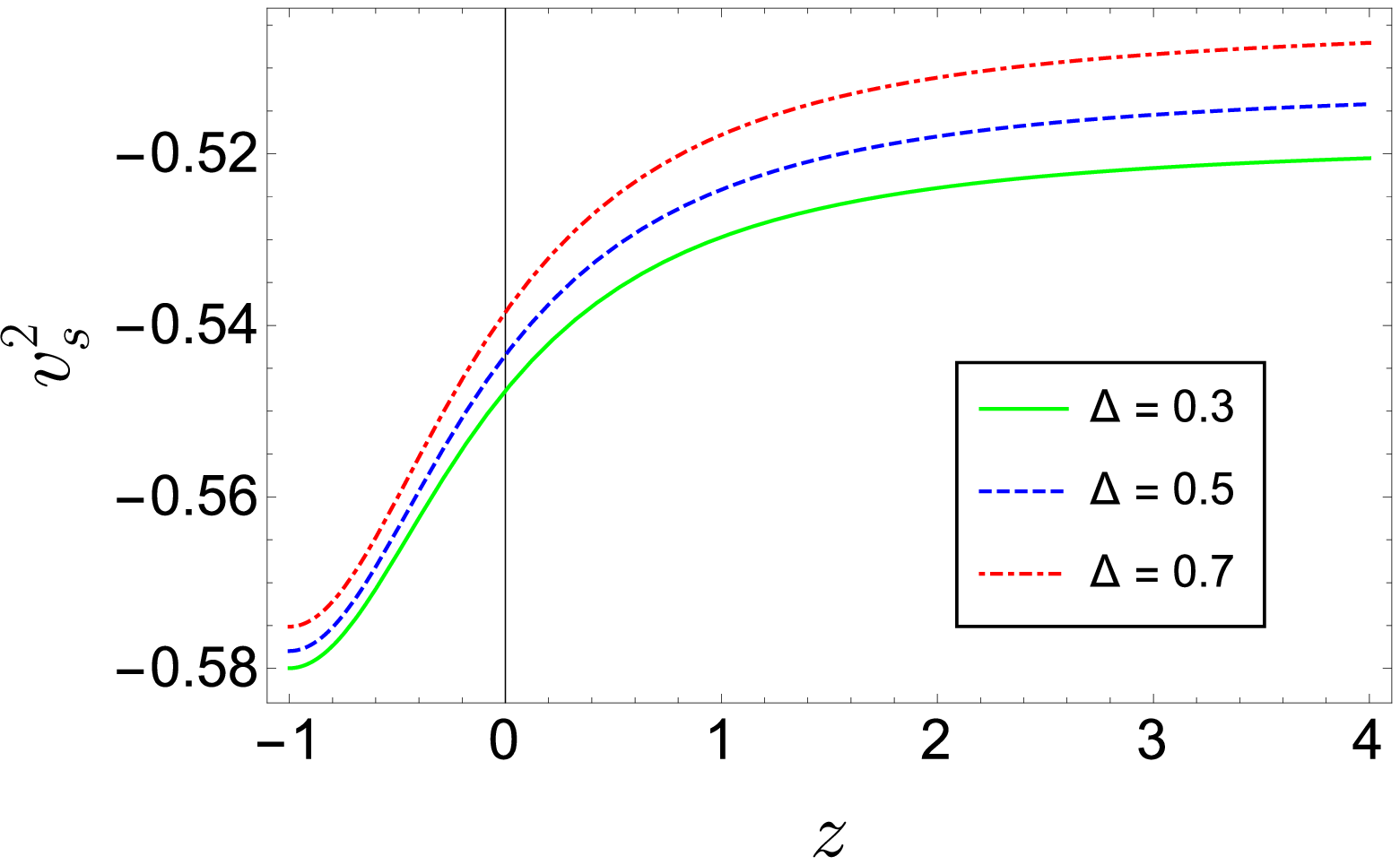}
\end{center}
\begin{center}
\includegraphics[width=8.5cm]{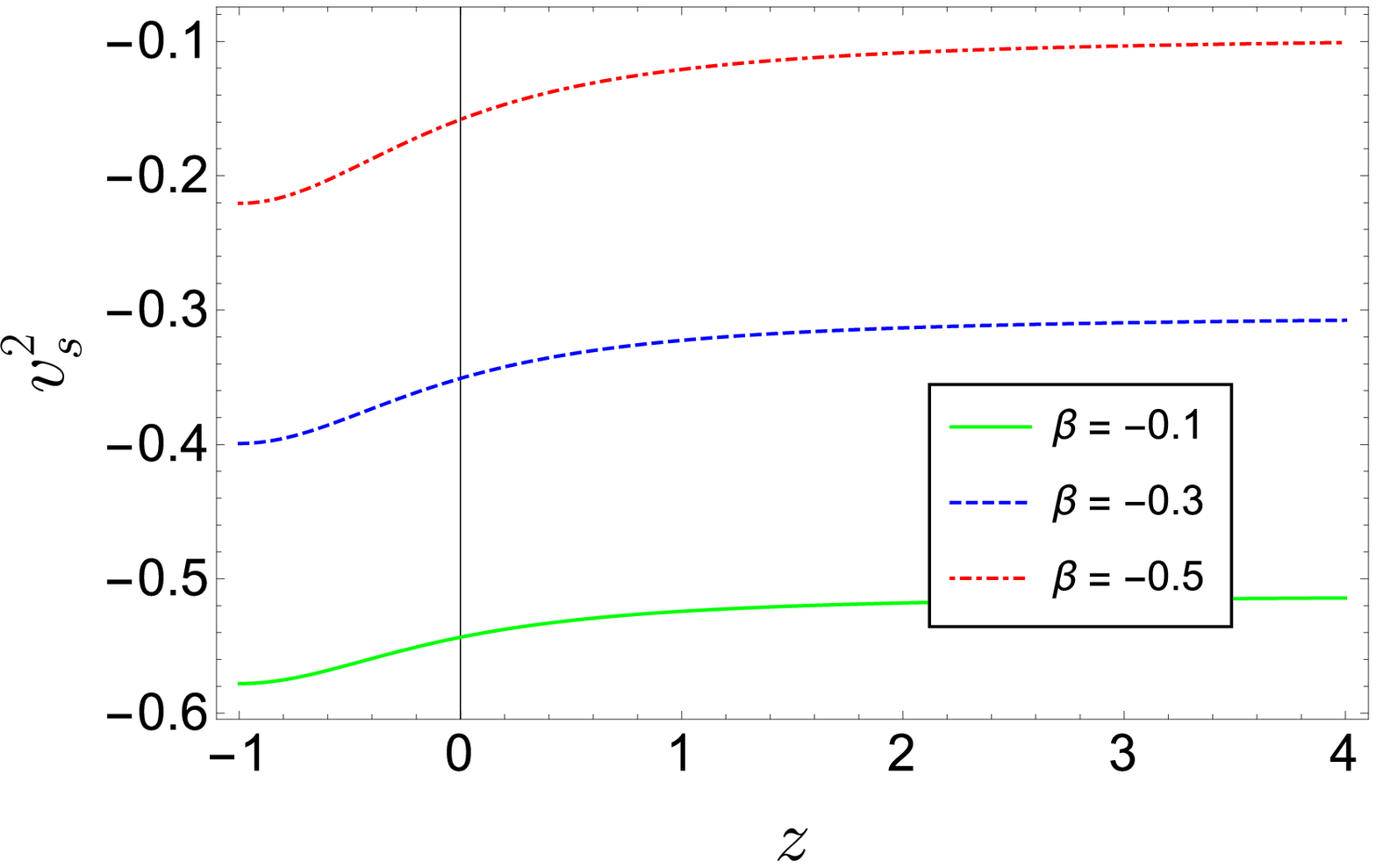}
\caption{Evolution of squared
speed of sound for different values
of $\alpha_0$ (upper panel), $\Delta$ (middle panel) and $\beta$ (lower panel) in interacting model. For all model parameters, we have set the same values as in Fig.~\ref{Fig1}. For the upper and middle panels, we have considered $\beta=-0.1$ (online colors).}
\label{Fig11}
\end{center}
\end{figure}

Let us finally consider how the  
classical stability is affected by Eq.~\eqref{inter}. After some algebra, 
we find the following expression for the squared sound speed
\begin{widetext}
\begin{eqnarray}
\nonumber
v_s^2&=&-1+C\alpha_0\,\alpha_1\,\alpha_2\left(\Delta-2\right)e^{\alpha_0t}H^{1-\Delta}\,+\,
\frac{2\,\alpha_1^{2\Delta-5}}{3C}\bigg\{1-\alpha_0\alpha_1\left(k-1\right)e^{\alpha_0t}\left[\frac{\alpha_1(k+2)}{\alpha_0}\,e^{\alpha_0t}+\alpha_2\left(k+2\right)\right]^{-\frac{k}{2+k}}\bigg\}\\[2mm]
\nonumber
&&\hspace{-9mm}\times\,\left[\frac{\alpha_1(k+2)}{\alpha_0}\,e^{\alpha_0t}+\alpha_2\left(k+2\right)\right]^{-\frac{4+k}{2+k}}\,H^{3-\Delta}\,+\,
C\alpha_0\,\alpha_1\,\alpha_2\,\left(\Delta-2\right)e^{\alpha_0t}H^{-\left(1+\Delta\right)}\left[\left(1-\Delta\right)\dot H+\alpha_0\,H\right]\frac{H^{1-\Delta}}{(2-\Delta)\dot H}\\[2mm]
\nonumber
&&\hspace{-9mm}+\,\frac{2\,\alpha_1^{2\Delta-5}H^{4-\Delta}}{3C\left(2-\Delta\right)\dot H}\left[\frac{\alpha_1(k+2)}{\alpha_0}\,e^{\alpha_0t}+\alpha_2\left(k+2\right)\right]^{-\frac{4+k}{2+k}}
\Bigg\{\bigg\{1-\alpha_0\,\alpha_1\left(k-1\right)e^{\alpha_0t}
\left[\frac{\alpha_1(k+2)}{\alpha_0}\,e^{\alpha_0t}+\alpha_2\left(k+2\right)\right]^{-\frac{k}{2+k}}\bigg\}\\[2mm]
\nonumber
&&\hspace{-9mm}\times\,\bigg\{\left(3-\Delta\right)\frac{\dot H}{H}-\left(\frac{4+k}{2+k}\right)
\left[\frac{\alpha_1(k+2)}{\alpha_0}\,e^{\alpha_0t}+\alpha_2\left(k+2\right)\right]^{-1}\bigg\}+\alpha_0^2\,\alpha_1\left(1-k\right)e^{\alpha_0t}
\bigg\{
\left[\frac{\alpha_1(k+2)}{\alpha_0}\,e^{\alpha_0t}+\alpha_2\left(k+2\right)\right]^{-\frac{k}{2+k}}\\[2mm]
&&\hspace{-9mm}-\,\frac{\alpha_1\,k}{\alpha_0}\,e^{\alpha_0t}
\left[\frac{\alpha_1(k+2)}{\alpha_0}\,e^{\alpha_0t}+\alpha_2\left(k+2\right)\right]^{-\frac{k+1}{2+k}}\bigg\}
\Bigg\}\,-\,\frac{\beta\dot q}{H}\,.
\end{eqnarray}
\end{widetext}

The behavior of $v_s^2$ is plotted in Fig.~\ref{Fig11}, 
indicating that increasing interactions might work in favor of a 
classical stabilization of  the model against small perturbations
(see the lower panel of Fig.~\ref{Fig11}, where it is shown that the larger the magnitude of $\beta$, the less stable
BHDE becomes). 
More discussion on the above results can be found in the last Section, 
along with further directions to explore.

\section{Observational constraints}
\label{Obco}

\subsection{Hubble's parameter}
{In order to explore the experimental
consistency of reconstructed BHDE and constrain Barrow exponent $\Delta$, let us study the evolution of Hubble's parameter $H(z)$ 
from Eq.~\eqref{qeq}. We develop our analysis for the non-interacting model and fixed parameters as in Sec.~\ref{reco}, but similar considerations can be extended to the case
when Eq.~\eqref{inter} is taken into account. Following~\cite{TeleBar}, 
we use data points obtained from 57 Hubble's parameter measurements in the range $0.07 \le z \le 2.36$. These points have been obtained through Differential Age (31 points), BAO and other methods (the remaining 26 points) and are outlined in Tab.~\ref{TabI} (see also~\cite{TeleBar} for more details).}

\begin{table}[h!]
  \centering
    \begin{tabular}{|c|c|c|c|c|c|}
    \hline
  \,  $z$\, &\, $H(z)\,$\, &\, $\sigma_H$\, & \,$z\,$ & \,$H(z)$\,& \, $\sigma_H$\,\\
  \hline
  \hline
  \,0.070\, &\, 69.0\, &\, 19.6\, &\, 0.4783\, &\hspace{-0.4mm} 80\, &\hspace{-0.2mm} 99\,\\
  \hline
0.90 & 69 & 12 & 0.480 & 97 & 62 \\
\hline
0.120 & 68.6 & 26.2 & 0.593 & 104 & 13 \\
\hline
0.170 & 83 & 8 & 0.6797 & 92 & 8\\
\hline
0.1791 & 75 & 4 & 0.7812 & 105 & 12\\
\hline
0.1993 & 75 & 5 & 0.8754 & 125 & 17\\
\hline
0.200 & 72.9 & 29.6 & 0.880 & 90 & 40\\
\hline
0.270 & 77 & 14 & 0.900 & 117 & 23\\
\hline
0.280 & 88.8 & 36.6 & 1.037 & 154 & 20\\
\hline
0.3519 & 83 & 14 & 1.300 & 168 & 17 \\
\hline
0.3802 & 83.0 & 13.5 & 1.363 & 160.0 & 33.6\\
\hline
0.400 & 95 & 17 & 1.430 & 177 & 18\\
\hline
0.4004 & 77.0 & 10.2 & 1.530 & 140 & 14\\
	\hline
0.4247 & 87.1 & 11.2 & 1.750 & 202 & 40\\
\hline
0.4497 & 92.8 & 12.9 & 1.965 & 186.5 & 50.4\\
\hline
0.470 & 89 & 34 & & & \\
\hline
      \end{tabular}
      
      \vspace{4mm}
      \begin{tabular}{|c|c|c|c|c|c|}
      \hline
  \,  $z$\, &\, $H(z)\,$\, &\, $\sigma_H$\, & \,$z\,$ & \,$H(z)$\,& \, $\sigma_H$\,\\
  \hline
  \hline
  \,0.24\, &\, 79.69\, &\, 2.99\, &\, 0.52\, &\, 94.35\, &\, 2.64\,\\
  \hline
0.30 &\, 81.70\, & 6.22 & 0.56 & 93.34 & 2.30\\
\hline
0.31 & 78.18 & 4.74 & 0.57 & 87.6 & 7.8\\
\hline
0.34 & 83.80 & 3.66 & 0.57 & 96.8 & 3.4\\
\hline
0.35 & 82.7 & 9.1 & 0.59 & 98.48 & 3.18\\
\hline
0.36 & 79.94 & 3.38 & 0.60 & 87.9 & 6.1\\
\hline
0.38 & 81.5 & 1.9 & 0.61 & 97.3 & 2.1\\
\hline
0.40 & 82.04 & 2.03 & 0.64 & 98.82 & 2.98\\
\hline
0.43 & 86.45 & 3.97 & 0.73 & 97.3 & 7.0\\
\hline
0.44& 82.6& 7.8& 2.30& 224.0& 8.6\\
\hline
0.44& 84.81& 1.83& 2.33& 224& 8\\
\hline
0.48 & 87.90 & 2.03 & 2.34 & 222.0 & 8.5\\
\hline
0.51 & 90.4 & 1.9 & 2.36 & 226.0 & 9.3\\
\hline
      \end{tabular}
  \caption{57 experimental points of $H(z)$ ($H$ is expressed in $\mathrm{km\,s^{-1}\,Mpc^{-1}}$ and $\sigma_H$ represents the uncertainty for each data point).}
  \label{TabI}
\end{table}

\begin{figure}[t]
\begin{center}
\includegraphics[width=8.5cm]{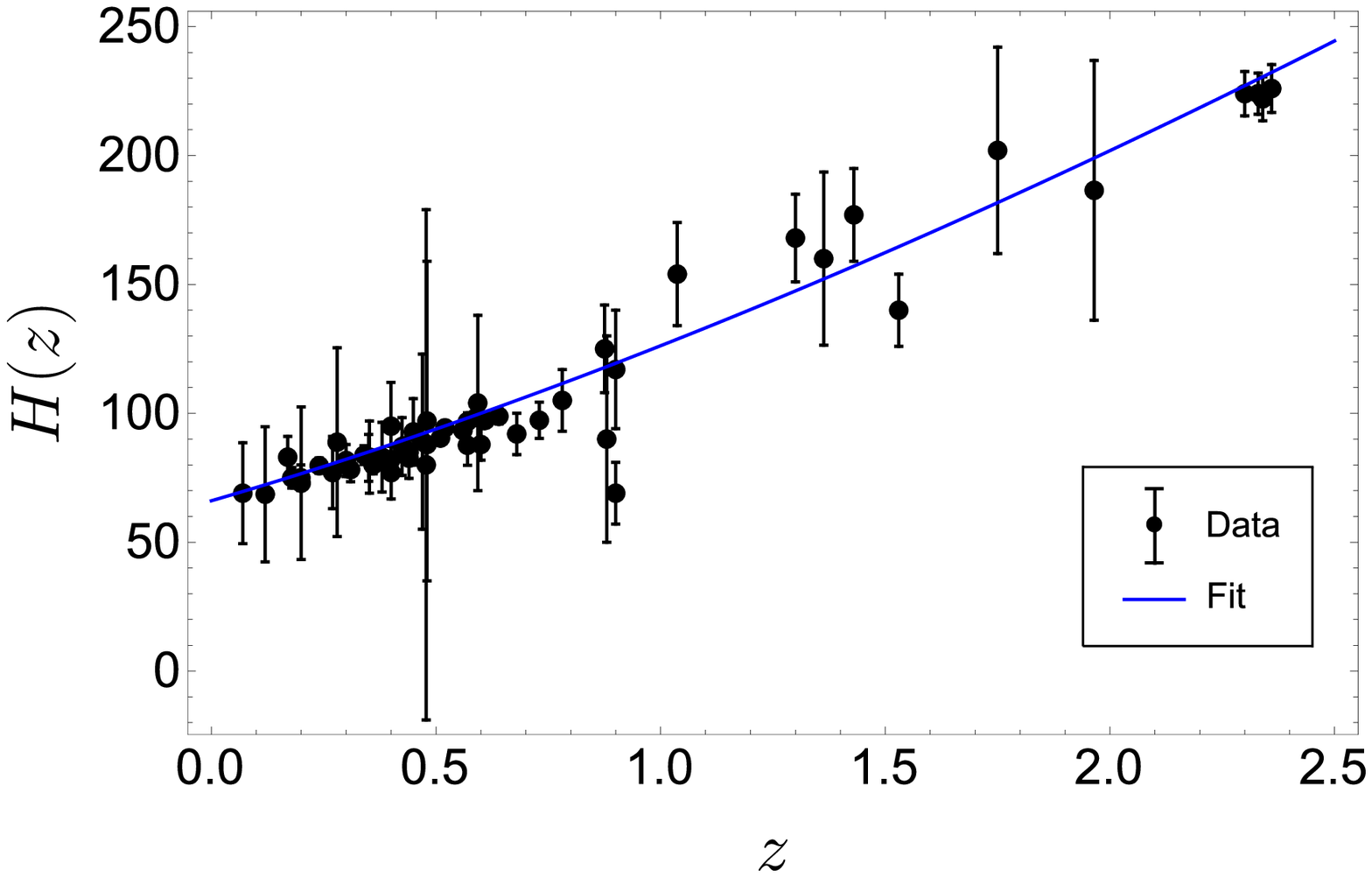}
\caption{Best fit curve of Hubble's parameter $H$ versus $z$. Dots represent data in Tab.~\ref{TabI}, while the curve is the theoretical fit (online colors).}
\label{PlotH}
\end{center}
\end{figure}

{The best fit is found by employing the statistical $R^2$-test, which is defined by 
\be
R^2\,=\,1-\frac{\sum_{i=1}^{57}\left[(H_i)_{ob}-(H_i)_{th}\right]^2}{\sum_{i=1}^{57}\left[(H_i)_{ob}-(H_i)_{mean}\right]^2}\,,
\ee
where $(H_i)_{ob}$ and $(H_i)_{th}$ are the observed and predicted
values of Hubble's parameter, respectively. 
Minimizing the departure of $|R^2-1|$ from zero gives the constraint 
$\Delta\simeq0.07$ at $95\%$ confidence level. This estimate
is in line with recent predictions in literature~\cite{Anagnostopoulos:2020ctz}, though being less stringent than that obtained 
via Big Bang Nucleosynthesis measurements~\cite{Barrow:2020kug}.
The fit in Fig.~\ref{PlotH} also allows us to
infer $H_0=(65.02\pm4.3)\,\mathrm{km\,s^{-1}\,Mpc^{-1}}$, which is close
to the recent observation from Planck Collaboration $H_0=\left(67.27\pm0.60\right)\mathrm{km\,s^{-1}\,Mpc^{-1}}$~\cite{Planck}, 
but deviates from $H_0=\left(74.03\pm1.42\right)\mathrm{km\,s^{-1}\,Mpc^{-1}}$ derived from 
2019 SH0ES collaboration~\cite{SH0ES}. }

\subsection{Cosmological perturbations in BHDE}
\label{CP}

{We now preliminarily investigate cosmological perturbations and structure formation in BHDE. For this analysis, we
are inspired by~\cite{Dago}. In particular, we work
in the linear regime on sub-horizon scales, 
studying the growth rate of matter fluctuations for clustering dark matter and a homogeneous dark energy component. More discussion along this line can be found in~\cite{Sheykhi:2022gzb}.
For the case of weakly interacting
dark components and scalar fluctuations of the metric
in the Newtonian gauge, the line element~\eqref{SBmetric} is modified 
by including a potential term $\phi$ as in~\cite{Mukha}. By introducing the density contrasts $\delta_i\equiv{\delta\rho_i}/{\rho_i}$ and divergences of the fluid velocities $\theta_i\equiv\vec{\nabla}\cdot\vec{v}_i$ for dark energy and dark matter, the evolution equations for the perturbations in the Fourier space take the form given in~\cite{Abramo}. We notice that DE effects on the growth of perturbations are only appreciable for $c^2_{eff}\equiv\delta p_{D}/\delta {\rho}_D\ll1$, since in this case dark energy and dark matter cluster in a similar way (adiabaticity condition)~\cite{Eric}. On the other hand, $c^2_{eff}\simeq1$ implies no growth because fluctuations would be suppressed by pressure. We can use evolution equations along with the relation $\frac{d}{dt}=\dot a\frac{d}{da}=aH\frac{d}{da}$ to extract differential equations for dark energy and matter perturbations in the form\footnote{Here we neglect anisotropic effects  because of some technicalities when solving evolution differential equations. This aspect will be considered in more detail in future work.}
\begin{eqnarray}
\label{31}
\frac{d^2\delta_m}{da^2}+A_m\frac{d\delta_m}{da}+B_m\delta_m&=&S_m\,,\\[2mm]
\label{32}
\frac{d^2\delta_D}{da^2}+A_D\frac{d\delta_D}{da}+B_D\delta_D&=&S_D\,,
\end{eqnarray}
where
\begin{eqnarray}
A_{m}&=&\frac{3}{2a}\left(	1-\omega_D\Omega_D\right)\,,\\[2mm]
B_{m}&=&0\,,\\[2mm]
S_m&=&\frac{3}{2a^2}\left[\Omega_m\delta_m+\left(1+3c_{eff}^2\right)\Omega_D\delta_D\right],\\[2mm]
A_{D}&=&\frac{1}{a}\left[\frac{3}{2}\left(1-\omega_D\Omega_D\right)-\frac{a}{1+\omega_D}\frac{d\omega_D}{da}-3\omega_D\right],\\[2mm]
\nonumber
B_{D}&=&\frac{1}{a^2}\left[3\left(\frac{1}{2}-\frac{3}{2}\omega_D\Omega_D-\frac{a}{1+\omega_D}\frac{d\omega_D}{da}-3c^2_{eff}\right)\right.\\[2mm]
&&\left.\times\,\left(c^2_{eff}-\omega_D\right)-3a\frac{d\omega_D}{da}+\frac{k}{a^2H^2}c^2_{eff}\right],\\[2mm]
S_D&=&\frac{3}{2a^2}\left(1+\omega_D\right)\left[\Omega_m\delta_m+\left(1+3c_{eff}^2\right)\Omega_D\delta_D\right],
\end{eqnarray}
where $\Omega_m=\rho_m/\rho_c$ and $\Omega_D=\rho_D/\rho_c$ 
are the fractional energy density parameters of dark matter and BHDE, 
respectively, and $\rho_c=3m_p^2H^2$ the critical energy density.
Hhere, $k$ identifies the (sub-horizon) scale in the Fourier space. }

{
To solve the above equations, we need to impose 
initial conditions. Concerning $\delta_m$, perturbed
Einstein equations lead to~\cite{Dago}
\begin{eqnarray}
\delta^{(in)}_m&=&-2\phi_{(in)}\left(1+\frac{k^2}{3a^2_{(in)}H^2_{(in)}}\right),\\[2mm]
\left(\frac{d\delta_m}{da}\right)^{(in)}&=&-\frac{2}{3}\frac{k^2\,\phi_{(in)}}{a^2_{(in)}H^2_{(in)}}\,.
\end{eqnarray}
Similarly, the adiabaticity condition gives for $\delta_D$~\cite{Kodama}
\begin{eqnarray}
\delta_D^{(in)}&=&\left(1+\omega_D\right)\delta_m^{(in)}\,,\\[2mm]
\left(\frac{d\delta_D}{da}\right)^{(in)}&=&\left(1+\omega_D\right)\left(\frac{d\delta_m}{da}\right)^{(in)}+\frac{d\omega_D}{da}\delta_m^{(in)}\,. 
\end{eqnarray}
For homogeneous BHDE (i.e. $\delta_D=0$),
the following evolution for the linear matter
perturbation on sub-horizon scales is obtained
\be
\label{dm}
\frac{d^2\delta_m}{da^2}+\left(\frac{3}{a}+\frac{1}{E}\frac{dE}{da}\right)\frac{d\delta_m}{da}-\frac{3\left(1-\Omega_D\right)}{2a^2}\delta_m\,=\,0\,,
\ee
where $E(a)$ is defined by
\be
E(a)\equiv \frac{H(a)}{H_0}\,.
\ee
Now, from Eq.~\eqref{dm} we can calculate 
the growth rate of matter density perturbations as
\be
f(a)\,\equiv\,a\,\frac{d\delta_m(a)}{da}\,.
\ee
We compare the above function with measurements of  $f \sigma_8(a)\equiv f(a)\sigma_8(a)$ from redshift-space distortion observations for $0<z<1.5$~\cite{Boss}, where $\sigma_8(a)=\sigma_8\delta_m(a)/\delta_m(1)$ is the linear-density field fluctuations in 
$8h^{-1}\,\mathrm{Mpc}$ radius and $\sigma_8$ its current value. 
We use the Gold-2017 dataset of 18 measurements of $f\sigma_8$~\cite{Nesseris},  which are rescaled with respect to a given fiducial cosmological model. We consider $\Lambda$CDM as fiducial framework
and introduce
\be
r(z)\,=\,\frac{H(z)\hspace{0.2mm}d_A(z)}{H_{fid}(z)d_{A,fid}(z)}\,,
\ee
where 
\be
H_{fid}(z)\,=\,H_0\sqrt{\Omega_{m0}\left(1+z\right)^3+\left(1-\Omega_{m0}\right)}\,,
\ee
is the Hubble expansion rate in $\Lambda$CDM model.  Measurements
can then be corrected by means of the vector
\be
\textbf{Y}\,=\,r(z_i)\hspace{0.2mm}f\hspace{0.2mm}\sigma_8^{obs}(z_i)-f\hspace{0.2mm}\sigma_8^{th}(z_i)\,.
\ee
Thus, the likelihood function is
given by $\mathcal{L}\,\propto\, e^{-\frac{1}{2}\hspace{0.2mm}\textbf{X}_{GRF}}$, where $\textbf{X}_{GRF}\,\equiv\,\textbf{Y}^T\,\textbf{C}_{GRF}^{-1}\,\textbf{Y}$ and $\textbf{C}_{GRF}$ is the covariance matrix of data listed in Tab.~\ref{TableII}.}

\begin{table}[t]
  \centering
    \begin{tabular}{|c|c|c|c|c|}
    \hline
  $z$\, &\, $f\sigma_8\,$\, &\, $z$\, &\, $f\sigma_8\,$\,\\
  \hline
  \hline
  \,\,0.02\,\, &\, 0.428\,$\pm$\,0.0465\, & \,\, 0.37\,\,&\, 0.4602\,$\pm$\,0.0378\, \\
  \hline
   \,0.02\, &\, 0.398\,$\pm$\,0.065\, & \, 0.32\,&\, 0.384\,$\pm$\,0.095\, \\
  \hline
   \,0.02\, &\, 0.314\,$\pm$\,0.048\, & \, 0.59\,&\, 0.488\,$\pm$\,0.060\, \\
  \hline
   \,0.10\, &\, 0.370\,$\pm$\,0.130\, & \, 0.44\,&\, 0.413\,$\pm$\,0.080\, \\
\hline
 \,0.15\, &\, 0.490\,$\pm$\,0.145\, & \, 0.60\,&\, 0.390\,$\pm$\,0.063\, \\
\hline
 \,0.17\, &\, 0.510\,$\pm$\,0.060\, & \, 0.73\,&\, 0.437\,$\pm$\,0.072\, \\
\hline
 \,0.18\, &\, 0.360\,$\pm$\,0.090\, & \, 0.60\,&\, 0.550\,$\pm$\,0.120\, \\
\hline
 \,0.38\, &\, 0.440\,$\pm$\,0.060\, & \, 0.86\,&\, 0.400\,$\pm$\,0.110\, \\
\hline
 \,0.25\, &\, 0.3512\,$\pm$\,0.0583\, & \, 1.40\,&\, 0.482\,$\pm$\,0.116\, \\
\hline
      \end{tabular}
      \caption{Dataset of 18 $f\sigma_8(z)$ measurements from different surveys~\cite{Nesseris}.}
      \label{TableII}
      \end{table}

\begin{figure}[t]
\begin{center}
\hspace{-1mm}\includegraphics[width=8.7cm]{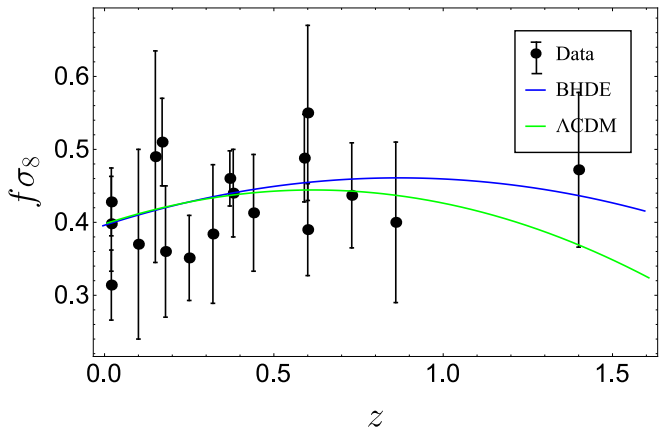}
\caption{Growth of matter perturbations in BHDE (blue curve) and $\Lambda$CDM (green curve). We have set $\Delta=0.07$ and $\sigma_8=0.895$, $\delta^{(in)}_m=10^{-2}$, $(d\delta_m/da)^{(in)}=1$ as initial conditions~\cite{Dago}, while all other relevant parameters have been fixed as in Fig.~\ref{Fig1} (online colors).}
 \label{Comp}
\end{center}
\end{figure}

{In Fig.~\ref{Comp} we show the 
growth rate of matter fluctuations for the
BHDE (blue curve) model compared to the $\Lambda$CDM scenario (green curve). We see that the BHDE curve is nearly
overlapped with the $\Lambda$CDM one
at low redshifts. On the other hand, the two plots depart as $z$ increases, 
with the BHDE curve fitting high-redshift data better than $\Lambda$CDM.
In particular,  perturbations are found to grow up faster compared to predictions from the standard Cosmology. 
Such a discrepancy at high redshift can be understood by stressing that
Barrow entropy is an effort to account for quantum-gravity
corrections on the horizon surface. Clearly, these effects
are expected to have more impact in the early Universe, when
gravity should be quantum. A similar result
has been recently exhibited in~\cite{Sheykhi:2022gzb}, where
the faster growth of perturbations has been attributed to the 
exquisitely fractal structure of Universe horizon in Barrow framework,
which can potentially  work in favor of the 
growth of fluctuations of energy density. 
}

A final comment is in order here: to show the viability  of a model, one needs complementary constraints by combing different observables. Indeed, single observational test cannot give solid sound result in principle. In this regard, we emphasize that 
cosmological models attempting to include quantum gravity effects in the standard Cosmology - like BHDE - are quite tough to test experimentally at present time, since quantum gravity effects (and their implications on the history of the Universe) are expected to be prominent in the very early Universe, where observational data are still lacking or not accurate enough. And in fact, constraints on Barrow entropy from relatively recent astrophysical/cosmological phenomena give very tiny deviations of Barrow parameter from zero, which in turn signal very small departure (if any) from Boltzmann-Gibbs entropy at that age (see, for instance,~\cite{Bar6,Anagnostopoulos:2020ctz,Barrow:2020kug,LucianoInf}). 
In this sense, and also motivated by the analysis of~\cite{Sheykhi:2022gzb}, we have then 
realized that a useful test bench for Barrow proposal could be the study of growth rate factor data of matter fluctuations and structure formation, where even tiny gravitational effects might appear amplified as a result of the amplification of primordial density fluctuations. To further support the viability of this study, we remark that a similar constraint based on the growth of perturbations has been proposed in~\cite{Dago} for the case of Tsallis Holographic Dark Energy. Clearly, one can still insist on parallel tests of Barrow model with different observables. We stress that investigation along this direction is very active in recent literature~\cite{Bar6,Anagnostopoulos:2020ctz,Barrow:2020kug,LucianoInf,Lucianoarx,Bar7}.

\section{Conclusions and Outlook}
\label{C&O}

\begin{center}
\begin{table*}[t]
    \begin{tabular}{|m{1.5cm}|m{4cm}|m{4.7cm}|m{5cm}|}
    \hline
 &\,\,\hspace{9mm}Free parameters\,\,&\,\, \hspace{7.5mm}Non-Interacting Model \,\, &\,\, \hspace{12.2mm}Observational value\,\, \\
 \hline
 \hline
\,\,$\hspace{4.8mm}\omega_{D_0}$ \,\, &\,\, \hspace{11mm} (See Fig.~\ref{Fig1}) \,\, &\,\, $ \hspace{14.5mm} [-1.07, -0.99] $ &\,\, \hspace{13.7mm}$[-1.38,-0.89]$\hspace{0.3mm}\cite{Planck} \,\, \\
           \hline
     \,\,  \hspace{4.8mm}    $q_0$  \,\, & \,\,  \hspace{18mm} \scriptsize{//} \,\, &\,\, $ \hspace{14.5mm} [-1.35,-1.23] $ &\,\, \hspace{13.7mm}$ [-1.37,-0.79]$\hspace{0.3mm}\cite{Camerana}\, \,\,\\
           \hline
        \,\,  \hspace{5.1mm}  $j$  \,\, &  \,\, \hspace{18mm}  \scriptsize{//} \,\, &\,\,\hspace{0.3mm} $j>0$ (for all $z$), $j(z\rightarrow-1)\rightarrow1$  &\,\, \hspace{16mm}$j=1$ ($\Lambda$CDM) \,\,\\
            \hline
            \,\,  \hspace{4.8mm}  $v_s^2$  \,\, & \,\, \hspace{17.7mm}  \scriptsize{//} \,\, &\,\, \hspace{5.5mm} $v_s^2<0$ (for $\alpha_0=1.1,1.3$) \par \vspace{1mm} \hspace{8mm} $v_s^2>0$ (for $\alpha_0=1.5$) \,\,&\,\, \hspace{25.4mm}- \,\,\\
            \hline
                    \,\,    $\omega_{GT}-\omega'_{GT}$  \,\, & \,\, \hspace{17.7mm}  \scriptsize{//} \,\,  &\,\,  \hspace{17.7mm} freezing \,\,&\,\, \hspace{25.4mm}- \,\,\\
            \hline
                  \,\,    \hspace{4.2mm} $H_0$  \,\, & \,\, \hspace{17.7mm}  \scriptsize{//} \,\,  &\,\,  \hspace{5mm} $(65.02\pm4.3)\,\mathrm{km\,s^{-1}\,Mpc^{-1}}$  \,\,&\,\hspace{1mm} $\left(67.27\pm0.60\right)\mathrm{km\,s^{-1}\,Mpc^{-1}}$~\cite{Planck}  \,\,\\
            \hline
    \end{tabular}
  \caption{Theoretical and observational values of EoS parameter, deceleration parameter, jerk parameter and squared sound speed for the best combination of arbitrary parameters of BHDE in SBT (for the jerk parameter we have considered the $\Lambda$CDM prediction as reference value).}
  \label{TabIII}
\end{table*}
\end{center}

In this work we have proposed a reconstruction
of Barrow Holographic Dark Energy in Saez-Ballester theory of gravity.
Motivated by recent observations from COBE, WMAP and Planck~\cite{Ani0,Ani1,Ani}, we have considered a Kantowski-Sachs Universe filled with dark matter and anisotropic BHDE as a background. To the best of our knowledge, this is 
the first time that BHDE is reconstructed in an anisotropic background like Kantowski-Sachs Universe.  Indeed, all previous studies are framed in the standard homogeneous and isotropic FRW Universe. In this sense, all the results obtained in the present analysis are novel, as they account for anisotropic effects in the evolution of the Universe.
By assuming the Hubble radius as an IR cutoff, we have
investigated both the cases of non-interacting and interacting
dark energy models, with special focus on the calculation
of skewness parameter, Equation-of-State parameter, 
deceleration parameter, jerk parameter and
squared sound speed. Among the main advantages  over other descriptions of dark energy, we have shown that our model correctly reproduces the
current accelerating phase of the cosmos, in contrast 
to standard HDE. Furthermore, the study of the squared speed of sound has revealed that our reconstruction is classically stable throughout the whole evolution of the Universe for higher skewness. By contrast, Saez Ballester-based reconstruction of Tsallis HDE, as well as non-interacting BHDE in Brans-Dicke Cosmology, are always unstable. Moreover, we have drawn 
the trajectories of $\omega_D-\omega'_D$ phase plane and discussed statefinder diagnosis for non-interacting mdoel. 
In order to constrain free parameters, we have
estimated current values of EoS parameter, jerk parameter
and deceleration parameter, and compared them with recent measurements from Planck+WP+BAO. We have finally
discussed the evolution of Hubble's parameter and 
growth rate of matter fluctuations in BHDE. 
Results are summarized in Tab.~\ref{TabIII}, showing that SBT-based reconstruction of non-interacting BHDE 
is observationally consistent for $1.1<\alpha_0<1.5$, 
with large values of $\alpha_0$ being compatible
with classical stability too. 
On the other hand, large (negative) interactions $\beta$
are phenomenologically disfavored, although they 
may contribute to stabilize the model
against small perturbations.  

\medskip 

Several aspects remain to be investigated: 

\begin{itemize}

\item[-]  as discussed in Sec.~\ref{reco}, our investigation
correctly explains the behavior of various model parameters
and the current accelerated expansion of the cosmos, 
though it does not predict its early-time decelerating phase. 
A possible explanation is that we are overlooking
some ordinary matter DoF, which mostly contributed to the energy content in the early Universe and caused its initial deceleration.
We reserve to improve our analysis in future investigation.

\item[-] In line with the study of~\cite{LucianoPRD,Mamon:2020spa,Chakraborty:2021uzp}, we aim at analyzing 
the thermodynamic implications of our model in
order to establish whether it is thermally stable.
This is essential to understand if SBT-based reconstruction 
of BHDE could clarify the yet unknown nature of DE.
In this regard, we remark that thermal stability has been studied in~\cite{LucianoPRD} by considering the heat capacities and compressibilities of both interacting and non-interacting BHDE. 
It has been shown that such a model does not satisfy the Thermal Stability Condition. Moreover, in~\cite{Lucianoarx} the generalized second law (GSL) of thermodynamics has been analyzed by also including radiation effects in the energy budget of the Universe. Since the total entropy variation is not necessarily a non-negative function, a violation of the GSL can potentially occur. We emphasize that these results are in line with the achievement of~\cite{Achiev} 
for the case of Tsallis Holographic Dark Energy and, more general, 
with the outcome of~\cite{Outc},  where it has been found that DE fluids with a time-dependent EoS parameter are in conflict with the physical constraints imposed by thermodynamics. All the above results have been obtained for the usual FRW background. It would be interesting to explore whether and how they appear in the case of anisotropic spacetime, such as Kantowski-Sachs Universe considered here. This aspect is under active investigation and will be presented elsewhere.

\item[-] In~\cite{Alleviate} Abdalla et al. have argued
that some HDE models can alleviate the $H_0$ tension 
since they predict $\omega_{D}<-1$, 
which seems a necessary condition to provide a solution based on
late-time modifications~\cite{LT}. This requirement 
is met in both the two models considered above, 
potentially giving some glimpses toward the resolution of the $H_0$
tension. On other hand, in~\cite{CapozLB} it has been argued that the Hubble tension could be somehow removed 
if the \emph{look-back time} is correctly referred to the redshift where the measurement is performed. This in turn rests upon the usage of the correct definition of $E(z)=H(z)/H_0$ in terms of the different contributions of radiation, matter, cosmological constant and spatial curvature to the density budget of the Universe. In this way, it has been shown that both values of the Hubble constant $H_0$ reported by the SH0ES and Planck collaborations~\cite{Alleviate,Khodadi} can be recovered. It is suggestive to explore whether a similar dynamical resolution of the Hubble tension can be obtained in the context of BHDE. Although we are here working with a fixed IR cutoff, we envisage that some sort of correspondence between BHDE and 
look-back time approach can be still established by allowing Barrow parameter $\Delta$ to be varying on time. Preliminary studies along this direction have been presented recently in~\cite{DiGennaro}. More work along this direction is inevitably needed.

\item[-]  BHDE is a generalization of standard Cosmology based on a deformation of the horizon entropy of the Universe. Extended cosmological scenarios, however, can also be obtained by modifying the geometric (i.e. gravitational)
sector of Einstein-Hilbert action or motivated by thermodynamic requirements over the cosmological kinematics. In these directions, interesting approaches are provided by the Extended Gravity Cosmography~\cite{EGC}, which is a model-independent framework to tackle the dark energy/modified gravity problem, and the thermodynamic parametrization of dark energy~\cite{TPDE}. Hence, a possible outlook is to study BHDE in parallel with such generalized approaches and possibly reinterpret Barrow's conjecture in this alternative language.

\item[-] In Sec.~\ref{CP} we have discussed cosmological 
perturbations and structure formation neglecting anisotropy of spacetime.
Clearly, a more comprehensive analysis requires 
including this feature too.

\item[-] A further challenging perspective is 
to extend the present analysis  to the
case of HDE based on Kaniadakis entropy~\cite{Kana1}
(Kaniadakis Holographic Dark Energy, KHDE),  
which is a self-consistent relativistic generalization of Boltzmann-Gibbs entropy parameterized by $-1<K<1$~\cite{GioKan,LucRev},
or HDE built on other commonly used entropies in physics, such
as Abe, Landsberg-Vedral, Sharma-Mittal and R\'eny entropies.
In this context, it would be interesting 
to find any connection between BHDE and such other models, and possibly constrain deformation parameters of these alternative modified entropies.

\item[-] A possible candidate for DE has been recently proposed in~\cite{Capol1,Capol2,Capol3} by looking at the properties
of the vacuum condensate of flavor mixed fields, in particular neutrinos~\cite{Blas95, Luc23}. Although the origin of this alternative explanation is rooted in particle physics, it would be interesting to discuss these results in connection with our model of HDE. This may also allows us to extend the paradigm of Barrow entropy to the framework of particle physics.

\item[-] Finally, since our model attempts to 
include quantum gravitational effects 
into HDE, it is important to analyze our results in connection
with predictions of more fundamental candidate theories  
of quantum gravity, such as String Theory, Loop Quantum Gravity
and Asymptotically Safe Gravity, or phenomenological approaches, such as Planck-scale deformations of the Heisenberg principle~\cite{KMM,Scardigli,LucGUP}. 
\end{itemize}

Work along these and other directions is still
in progress and results will be presented elsewhere.

\acknowledgments 
The author acknowledges S. Odintsov 
for comments on the original manuscript. He is also grateful
to the Spanish ``Ministerio de Universidades'' 
for the awarded Maria Zambrano fellowship and funding received
from the European Union - NextGenerationEU. 
He finally expresses his gratitude for participation in the COST
Action CA18108  ``Quantum Gravity Phenomenology in the Multimessenger Approach'' and LISA Cosmology Working group.


\begin{thebibliography}{99}
%
\bibitem{Primack:2006it}
J.~R.~Primack,
Nucl. Phys. B Proc. Suppl. \textbf{173}, 1 (2007). 
\bibitem{Supern}
A.~G.~Riess \textit{et al.} [Supernova Search Team],
Astron. J. \textbf{116}, 1009 (1998). 
%
\bibitem{Supernbis}
S.~Perlmutter \textit{et al.} [Supernova Cosmology Project],
Astrophys. J. \textbf{517}, 565 (1999).
%
\bibitem{Supernter}
D.~N.~Spergel \textit{et al.} [WMAP],
Astrophys. J. Suppl. \textbf{148}, 175 (2003).
%
\bibitem{Supernquar}
M.~Tegmark \textit{et al.} [SDSS],
Phys. Rev. D \textbf{69}, 103501 (2004).
%
\bibitem{Supernquin}
P.~A.~R.~Ade \textit{et al.} [Planck],
Astron. Astrophys. \textbf{571}, A16 (2014).
%
\bibitem{Vag1}
S.~Vagnozzi, L.~Visinelli, P.~Brax, A.~C.~Davis and J.~Sakstein,
Phys. Rev. D \textbf{104},  063023 (2021).
%
\bibitem{Vag2}
F.~Ferlito, S.~Vagnozzi, D.~F.~Mota and M.~Baldi,
Mon. Not. Roy. Astron. Soc. \textbf{512},  1885 (2022).
%
\bibitem{ER}
P.~Salucci, G.~Esposito, G.~Lambiase, E.~Battista, M.~Benetti, D.~Bini, L.~Boco, G.~Sharma, V.~Bozza and L.~Buoninfante, \textit{et al.}
Front. in Phys. \textbf{8}, 603190 (2021). 
%
\bibitem{CapozDela}
S.~Capozziello and M.~De Laurentis,
Phys. Rept. \textbf{509}, 167 (2011).
%
\bibitem{Cohen:1998zx}
A.~G.~Cohen, D.~B.~Kaplan and A.~E.~Nelson,
Phys. Rev. Lett. \textbf{82}, 4971 (1999).
%
\bibitem{Horava:2000tb}
P.~Horava and D.~Minic,
Phys. Rev. Lett. \textbf{85}, 1610 (2000).
%
\bibitem{Thomas:2002pq}
S.~D.~Thomas,
Phys. Rev. Lett. \textbf{89}, 081301 (2002).
%
\bibitem{Li:2004rb}
M.~Li,
Phys. Lett. B \textbf{603}, 1 (2004).
%
\bibitem{Hsu:2004ri}
S.~D.~H.~Hsu,
Phys. Lett. B \textbf{594}, 13 (2004).
%
\bibitem{Huang:2004ai}
Q.~G.~Huang and M.~Li,
JCAP \textbf{08}, 013 (2004).
%
\bibitem{Nojiri:2005pu}
S.~Nojiri and S.~D.~Odintsov,
Gen. Rel. Grav. \textbf{38}, 1285  (2006). 
%
\bibitem{Wang:2005ph}
B.~Wang, C.~Y.~Lin and E.~Abdalla,
Phys. Lett. B \textbf{637}, 357 (2006).
%
\bibitem{Setare:2006sv}
M.~R.~Setare,
Phys. Lett. B \textbf{642}, 421 (2006).
%
\bibitem{Guberina:2006qh}
B.~Guberina, R.~Horvat and H.~Nikolic,
JCAP \textbf{01}, 012 (2007).
%
\bibitem{Granda}
L.N. Granda, A. Oliveros, Phys. Lett. B \textbf{671275}, 199 (2009).
%
\bibitem{Sheykhi:2011cn}
A.~Sheykhi,
Phys. Rev. D \textbf{84}, 107302 (2011).
%
\bibitem{Bamba:2012cp}
K.~Bamba, S.~Capozziello, S.~Nojiri and S.~D.~Odintsov,
Astrophys. Space Sci. \textbf{342}, 155 (2012).
%
\bibitem{Ghaffari:2014pxa}
S.~Ghaffari, M.~H.~Dehghani and A.~Sheykhi,
Phys. Rev. D \textbf{89}, 123009 (2014).
%
\bibitem{Wang:2016och}
S.~Wang, Y.~Wang and M.~Li,
Phys. Rept. \textbf{696}, 1 (2017).
%
\bibitem{Odi1}
S.~Nojiri and S.~D.~Odintsov,
Eur. Phys. J. C \textbf{77} 528 (2017).
%
\bibitem{Moradpour:2020dfm}
H.~Moradpour, A.~H.~Ziaie and M.~Kord Zangeneh,
Eur. Phys. J. C \textbf{80}, 732 (2020).
%
\bibitem{Zhang}
X.~Zhang and F.~Q.~Wu,
Phys. Rev. D \textbf{72}, 043524 (2005).
%
\bibitem{Li}
M.~Li, X.~D.~Li, S.~Wang and X.~Zhang,
JCAP \textbf{06}, 036 (2009).
%
\bibitem{Zhang2}
X.~Zhang,
Phys. Rev. D \textbf{79}, 103509 (2009).
%
\bibitem{Lu}
J.~Lu, E.~N.~Saridakis, M.~R.~Setare and L.~Xu,
JCAP \textbf{03}, 031 (2010). 
%
\bibitem{Nojiri:2019kkp}
S.~Nojiri, S.~D.~Odintsov and E.~N.~Saridakis,
Phys. Lett. B \textbf{797} 134829  (2019). 
%
%
%
%
%
%
\bibitem{Odi}
S.~Nojiri, S.~D.~Odintsov and V.~Faraoni,
Phys. Rev. D \textbf{105} 044042  (2022). 
%
\bibitem{Odi2}
S.~Nojiri, S.~D.~Odintsov and T.~Paul,
Symmetry \textbf{13} 928 (2021).
%
\bibitem{Tsallis1}
M.~Tavayef, A.~Sheykhi, K.~Bamba and H.~Moradpour,
Phys. Lett. B \textbf{781}, 195 (2018).
%
\bibitem{Tsallis2}
E.~N.~Saridakis, K.~Bamba, R.~Myrzakulov and F.~K.~Anagnostopoulos,
JCAP \textbf{12}, 012 (2018). 
%
\bibitem{Tsallis3}
S.~Nojiri, S.~D.~Odintsov and E.~N.~Saridakis,
Eur. Phys. J. C \textbf{79}, 242 (2019)..
%
\bibitem{Tsallis4}
G.~G.~Luciano and J.~Gine,
Phys. Lett. B \textbf{833}, 137352 (2022).
%
\bibitem{Kana1}
N.~Drepanou, A.~Lymperis, E.~N.~Saridakis and K.~Yesmakhanova,
Eur. Phys. J. C \textbf{82}, 449 (2022).
%
\bibitem{Kana2}
A.~Hern\'andez-Almada, G.~Leon, J.~Maga\~na, M.~A.~Garc\'\i{}a-Aspeitia, V.~Motta, E.~N.~Saridakis and K.~Yesmakhanova,
Mon. Not. Roy. Astron. Soc. \textbf{511}, 4147 (2022).
%
\bibitem{Kana3}
G.~G.~Luciano,
Eur. Phys. J. C \textbf{82}, 314 (2022). 
%
\bibitem{Bar1}
E. N. Saridakis, Phys. Rev. D \textbf{102}, 123525 (2020).
%
\bibitem{Bar2}
M.~P.~Dabrowski and V.~Salzano,
Phys. Rev. D \textbf{102}, 064047 (2020).
%
\bibitem{Bar3}
A.~Sheykhi,
Phys. Rev. D \textbf{103},  123503 (2021).
%
\bibitem{Bar4}
P.~Adhikary, S.~Das, S.~Basilakos and E.~N.~Saridakis,
Phys. Rev. D \textbf{104}, 123519 (2021).
%
\bibitem{Bar5}
S.~Nojiri, S.~D.~Odintsov and T.~Paul,
Phys. Lett. B \textbf{825}, 136844 (2022). 
%
\bibitem{Bar6}
G.~G.~Luciano and E.~N.~Saridakis,
Eur. Phys. J. C \textbf{82}, 558 (2022).
%
\bibitem{Bar7}
S.~Ghaffari, G.~G.~Luciano and S.~Capozziello,
Eur. Phys. J. Plus \textbf{138}, 82 (2023).
%
\bibitem{Lucianoarx}
G.~G.~Luciano and J.~Gin\'e,
[arXiv:2210.09755 [gr-qc]].
%
\bibitem{BarUlt}
N.~Boulkaboul,
Phys. Dark Univ. \textbf{40}, 101205 (2023). 
%
\bibitem{EPL}
P.~S.~Ens and A.~F.~Santos,
EPL \textbf{131},  40007 (2020).
%
\bibitem{Chinese}
M.~Zubair and L.~R.~Durrani,
Chin. J. Phys. \textbf{69}, 153 (2021).
%
\bibitem{fgtgrav}
M.~Sharif and S.~Saba,
Symmetry \textbf{11}, 92 (2019). 
%
\bibitem{Wahe}
S.~Waheed,
Eur. Phys. J. Plus \textbf{135},  11 (2020). 
%
\bibitem{GhaffariBD}
S.~Ghaffari, H.~Moradpour, I.~P.~Lobo, J.~P.~Morais Gra\c{c}a and V.~B.~Bezerra,
Eur. Phys. J. C \textbf{78}, 706 (2018).
%
\bibitem{LogBD}
Y.~Aditya, S.~Mandal, P.~K.~Sahoo and D.~R.~K.~Reddy,
Eur. Phys. J. C \textbf{79}, 1020 (2019). 
%
\bibitem{LiuT}
Y.~Liu,
Eur. Phys. J. Plus \textbf{136}, 579 (2021). 
%
\bibitem{Sarkar}
A.~Sarkar and S.~Chattopadhyay,
Int. J. Geom. Meth. Mod. Phys. \textbf{18}, 2150148 (2021). 
%
\bibitem{LucianoPRD}
G.~G.~Luciano,
Phys. Rev. D \textbf{106}, 083530 (2022). 
%
\bibitem{TeleBar}
M.~Koussour, S.~H.~Shekh and M.~Bennai,
Int. J. Mod. Phys. A \textbf{37}, 2250184 (2022).
%
\bibitem{B2020}
J.~D.~Barrow,
Phys. Lett. B \textbf{808}, 135643 (2020).
%
\bibitem{SB}
D. Saez and V.J. Ballester, Phys. Lett. A {\bf 113}, 467 (1986). 
%
\bibitem{SBTBianchi1}
A.~Pradhan, A.~Kumar Singh and D.~S. Chouhan, 
Int. J. Theor. Phys. {\bf 52}, 266 (2013).
%
\bibitem{Rasouli1}
S.~M.~M.~Rasouli and P.~Vargas Moniz,
Class. Quant. Grav. \textbf{35}, 025004 (2018).
%
%
\bibitem{SBTBianchi2}
U.~K. Sharma, R.~Zia, A.~Pradhan, J. Astrophys. Astr. {\bf40}, 2 (2019). 
%
\bibitem{Rasouli2}
S.~M.~M.~Rasouli, R.~Pacheco, M.~Sakellariadou and P.~V.~Moniz,
Phys. Dark Univ. \textbf{27}, 100446 (2020).
%
\bibitem{Santhi}
Y. Sobhanbabu and M. Vijaya Santhi, 
Eur. Phys. J. C \textbf{81}, 1040 (2021).
%
\bibitem{Kim}
H.~Kim, Mon. Not. R. Astron. Soc. {\bf 364}, 813 (2005).
%
\bibitem{Guth}
A.~H.~Guth, Phys. Rev. D \textbf{23}, 347 (1981).
%
\bibitem{Linde}
A.~Linde, Phys. Lett. B \textbf{108}, 389 (1982).
%
\bibitem{KSM}
R. Kantowski and R. K Sachs, J. Math. Phys \textbf{7}, 3 (1966).
%
\bibitem{Req1}
A.~Sharma, K.~Banerjee and J.~Bhattacharyya,
Phys. Rev. D \textbf{106}, 063518 (2022).
%
\bibitem{Ani0}
W.~J.~Percival \textit{et al.} [2dFGRS],
Mon. Not. Roy. Astron. Soc. \textbf{327}, 1297 (2001).
%
\bibitem{Ani1}
A.~C.~Pope \textit{et al.} [SDSS],
Astrophys. J. \textbf{607}, 655 (2004).
%
\bibitem{Ani}
K.~Migkas, G.~Schellenberger, T.~H.~Reiprich, F.~Pacaud, M.~E.~Ramos-Ceja and L.~Lovisari,
Astron. Astrophys. \textbf{636}, A15 (2020). 
%
\bibitem{Bianchi_I}
B.~C.~Paul, B.~C.~Roy and A.~Saha,
Eur. Phys. J. C \textbf{82},  76 (2022).
%
\bibitem{Collins}
C.~B.~Collins, E.~N.~Glass and D.~A.~Wilkinson,
Gen. Rel. Grav. \textbf{12}, 805 (1980).
%
\bibitem{Skewness1}
K.S. Adhav, Int. J. Astron. Astrophys. {\bf 1}, 204 (2011).
%
\bibitem{Skewness2}
M.V. Santhi et al., Can. J. Phys. {\bf 95}, 179 (2017). 
%
\bibitem{Wang:2016lxa}
B.~Wang, E.~Abdalla, F.~Atrio-Barandela and D.~Pavon,
Rept. Prog. Phys. \textbf{79}, 096901 (2016).
%
\bibitem{Anagnostopoulos:2020ctz}
F.~K.~Anagnostopoulos, S.~Basilakos and E.~N.~Saridakis,
Eur. Phys. J. C \textbf{80}, 826 (2020).
%
\bibitem{Barrow:2020kug}
J.~D.~Barrow, S.~Basilakos and E.~N.~Saridakis,
Phys. Lett. B \textbf{815}, 136134 (2021).  
%
\bibitem{LucianoInf}
G.~G.~Luciano,
[arXiv:2301.12509 [gr-qc]].
%
\bibitem{Planck}
N.~Aghanim \textit{et al.} [Planck],
Astron. Astrophys. \textbf{641}, A6 (2020)
[erratum: Astron. Astrophys. \textbf{652}, C4 (2021)]. 
%
\bibitem{SH0ES}
 A.~G.~Riess, S.~Casertano, W.~Yuan, L.~M.~Macri and D.~Scolnic,
Astrophys. J. \textbf{876}, 85 (2019).
%
\bibitem{Dago}
R.~D'Agostino,
Phys. Rev. D \textbf{99}, 103524 (2019).
%
\bibitem{Sheykhi:2022gzb}
A.~Sheykhi and B.~Farsi,
Eur. Phys. J. C \textbf{82}, 1111 (2022). 
%
\bibitem{Mukha}
V. F. Mukhanov, H. A. Feldman, R. Brandenberger,
Phys. Rep. {\bf215}, 206 (1992). 
%
\bibitem{Abramo}
L.~R.~Abramo, R.~C.~Batista, L.~Liberato and R.~Rosenfeld,
Phys. Rev. D \textbf{79}, 023516 (2009).
%
\bibitem{Eric}
J.~K.~Erickson, R.~R.~Caldwell, P.~J.~Steinhardt, C.~Armendariz-Picon and V.~F.~Mukhanov,
Phys. Rev. Lett. \textbf{88}, 121301 (2002).
%
\bibitem{Kodama}
H.~Kodama and M.~Sasaki,
Prog. Theor. Phys. Suppl. \textbf{78}, 1 (1984).
%
\bibitem{Boss}
S.~Alam \textit{et al.} [BOSS],
Mon. Not. Roy. Astron. Soc. \textbf{470}, 2617 (2017).
%
\bibitem{Nesseris}
S. Nesseris, G. Pantazis, L. Perivolaropoulos, Phys. Rev.
D {\bf96}, 023542 (2017).
%
\bibitem{Raju}
K.~D.~Raju, M.~P.~V.~V.~Bhaskara Rao, Y.~Aditya, T.~Vinutha and D.~R.~K.~Reddy,
Can. J. Phys. \textbf{98}, 993 (2020).
%
\bibitem{CL}
R.~R.~Caldwell and E.~V.~Linder,
Phys. Rev. Lett. \textbf{95}, 141301 (2005). 
%
\bibitem{Sche}
R.~J.~Scherrer,
Phys. Rev. D \textbf{73}, 043502 (2006). 
%
\bibitem{Chiba}
T.~Chiba,
Phys. Rev. D \textbf{73}, 063501 (2006)
[erratum: Phys. Rev. D \textbf{80}, 129901 (2009)]. 
%
\bibitem{Guo}
Z.~K.~Guo, Y.~S.~Piao, X.~Zhang and Y.~Z.~Zhang,
Phys. Rev. D \textbf{74}, 127304 (2006). 
%
\bibitem{Sharif}
M.~Sharif and A.~Jawad,
Eur. Phys. J. C \textbf{72}, 2097 (2012). 
%
\bibitem{Camerana}
D.~Camarena and V.~Marra,
Phys. Rev. Res. \textbf{2}, 013028 (2020).
%
\bibitem{Blandford:2004ah}
R.~D.~Blandford, M.~A.~Amin, E.~A.~Baltz, K.~Mandel and P.~J.~Marshall,
ASP Conf. Ser. \textbf{339}, 27 (2005).
%
\bibitem{Pert1}
H.~Kim,
Mon. Not. Roy. Astron. Soc. \textbf{364}, 813 (2005).
%
\bibitem{rs}
V.~Sahni, T. D. Saini, A. A. Starobinsky and U. Alam, JETP Lett. {\bf 77}, 201 (2003). 
%
\bibitem{class}
U. Alam, V.~Sahni, T. D. Saini, A. A. Starobinsky, Mon. Not. Roy. Astron. Soc. \textbf{344}, 1057 (2003). 
%
\bibitem{ComT0}
H. Wei, Nucl. Phys. B \textbf{845}, 381 (2011). 
%
\bibitem{ComT1}
H.~Wei,
Commun. Theor. Phys. \textbf{56}, 972 (2011).
%
\bibitem{ComT2}
Y.~D.~Xu and Z.~G.~Huang,
Astrophys. Space Sci. \textbf{350}, 855 (2014).
%
\bibitem{Q1}
L.~P.~Chimento, A.~S.~Jakubi, D.~Pavon and W.~Zimdahl,
Phys. Rev. D \textbf{67}, 083513 (2003).
%
\bibitem{Q2}
D.~Pavon and W.~Zimdahl,
Phys. Lett. B \textbf{628}, 206 (2005).
%
\bibitem{Q3}
C.~G.~Boehmer, G.~Caldera-Cabral, R.~Lazkoz and R.~Maartens,
Phys. Rev. D \textbf{78}, 023505 (2008).
%
\bibitem{Q4}
G.~Caldera-Cabral, R.~Maartens and L.~A.~Urena-Lopez,
Phys. Rev. D \textbf{79}, 063518 (2009).
%
\bibitem{Mamon:2020spa}
A.~A.~Mamon, A.~Paliathanasis and S.~Saha,
Eur. Phys. J. Plus \textbf{136}, 134 (2021).
%
\bibitem{Chakraborty:2021uzp}
G.~Chakraborty, S.~Chattopadhyay, E.~G\"udekli and I.~Radinschi,
Symmetry \textbf{13}, 562 (2021).
%
\bibitem{Achiev}
M.~Abdollahi Zadeh, A.~Sheykhi and H.~Moradpour,
Gen. Rel. Grav. \textbf{51}, 12 (2019).
%
\bibitem{Outc}
E. M. Barboza, R. C. Nunes, E. M. C. Abreu and
J. A. Neto, Phys. Rev. D \textbf{92}, 083526 (2015). 
%
\bibitem{Alleviate}
E.~Abdalla, G.~Franco Abell\'an, A.~Aboubrahim, A.~Agnello, O.~Akarsu, Y.~Akrami, G.~Alestas, D.~Aloni, L.~Amendola and L.~A.~Anchordoqui, \textit{et al.}
JHEAp \textbf{34}, 49 (2022). 
%
\bibitem{LT}
S.~Vagnozzi,
Phys. Rev. D \textbf{102}, 023518 (2020). 
%
\bibitem{CapozLB}
S.~Capozziello, G.~Sarracino and A.~D.~A.~M.~Spallicci,
Phys. Dark Univ. \textbf{40}, 101201 (2023). 
%
\bibitem{Khodadi}
M.~Khodadi and M.~Schreck,
Phys. Dark Univ. \textbf{39}, 101170 (2023). 
%
\bibitem{DiGennaro}
S.~Di Gennaro and Y.~C.~Ong,
Universe \textbf{8}, 541 (2022). 
%
\bibitem{EGC}
S.~Capozziello, R.~D'Agostino and O.~Luongo,
Int. J. Mod. Phys. D \textbf{28}, 1930016 (2019).
%
\bibitem{TPDE}
S.~Capozziello, R.~D'Agostino and O.~Luongo,
Phys. Dark Univ. \textbf{36}, 101045 (2022). 
%
\bibitem{GioKan}
G. Kaniadakis, Phys. Rev. E {\bf 66}, 056125 (2002).
%
\bibitem{LucRev}
G.G. Luciano, Entropy {\bf 24}, 1712 (2022). 
%
\bibitem{Capol1}
A.~Capolupo and A.~Quaranta,
Phys. Lett. B \textbf{840}, 137889 (2023).
%
\bibitem{Capol2}
A.~Capolupo and A.~Quaranta,
Phys. Lett. B \textbf{839}, 137776 (2023)
%
\bibitem{Capol3}
A.~Capolupo,
Adv. High Energy Phys. \textbf{2018}, 9840351 (2018).
%
\bibitem{Blas95}
M.~Blasone and G.~Vitiello,
Annals Phys. \textbf{244}, 283 (1995).
%
\bibitem{Luc23}
G.~G.~Luciano,
Eur. Phys. J. Plus \textbf{138},, 83 (2023).
%
\bibitem{KMM}
A.~Kempf, G.~Mangano and R.~B.~Mann,
Phys. Rev. D \textbf{52}, 1108 (1995). 
%
\bibitem{Scardigli}
F.~Scardigli,
Phys. Lett. B \textbf{452}, 39 (1999).
%
\bibitem{LucGUP}
G.~G.~Luciano,
Eur. Phys. J. C \textbf{81}, 672 (2021). 



\end{thebibliography}
\end{document}